\documentclass[
aps,
prb,
reprint,
groupedaddress,
superscriptaddress,
amsmath,
amssymb
]{revtex4-2}

\usepackage{graphicx}
\usepackage{hyperref}
\usepackage{braket}
\usepackage{algorithmic}
\usepackage{mathtools}
\usepackage{amsmath,amssymb}
\usepackage[columnwise]{lineno}

\usepackage{amsthm}
\usepackage{ascmac}
\usepackage{amsfonts}
\usepackage{extarrows}
\usepackage{color}
\usepackage{bm}
\usepackage{ulem}
\newcommand{\I}{{\mathrm{i}}}

\allowdisplaybreaks[1]

\allowdisplaybreaks[4]

\begin{document}

\title{Photoemission Orbital Tomography Using Robust Sparse PhaseLift}
\author{Kaori Niki}
\affiliation{Graduate School of Science, Chiba University, Chiba, Japan}
\author{Rena Asano}
\affiliation{Graduate School of Science, Chiba University, Chiba, Japan}
\author{Ryuji Sakanoue}
\affiliation{Graduate School of Science, Chiba University, Chiba, Japan}
\author{Manabu Hagiwara}
\affiliation{Graduate School of Science, Chiba University, Chiba, Japan}
\author{Kazushi Mimura}
\affiliation{
School of Computing, Tokyo Institute of Technology, Yokohama, Kanagawa, Japan}
\affiliation{
Graduate School of Information Sciences, Hiroshima City University, Hiroshima, Japan}
\date{\today}

\begin{abstract}
\par
Photoemission orbital tomography (POT) from photoelectron momentum maps (PMMs) has enabled detailed analysis of the shape and energy of molecular orbitals in the adsorbed state. This study proposes a new POT method based on the PhaseLift. Molecular orbitals, including three-dimensional phases, can be identified from a single PMM by actively providing atomic positions and basis. Moreover, our method is robust to noise and can perfectly discriminate adsorption-induced molecular deformations with an accuracy of 0.05 [\AA]. Our new method enables simultaneous analysis of the three-dimensional shapes of molecules and molecular orbitals and thus paves the way for advanced quantum-mechanical interpretation of adsorption-induced electronic state changes and photo-excited inter-molecular interactions. 
\end{abstract}
\maketitle

\section{Introduction}
\label{sec:intro}
\par
Wavenumber-resolved photoelectron spectroscopy captures photoelectrons emitted in all directions with high speed and high precision. Therefore, it enables the analysis of the electronic structure of organic thin films without damaging the sample.Until now, changes in the charge distribution of adsorbed molecules have been discussed based on changes in the work function measured by capturing photoelectrons emitted in specific directions using X-ray photoelectron spectroscopy (XPS) or ultraviolet photoelectron spectroscopy (UPS) \cite{king, king_28, kera, king_33, F4TCNQ}. 
\par
However, analysis of photoelectron momentum maps (PMMs), which contain information on electronic states in all directions, is expected to reveal more physical properties. Derivation of the inverse Fourier transforms (FTs) of PMMs enables the detailed analysis of the shape and energy of molecular orbitals. This method is known as the photoemission orbital tomography (POT) \cite{Science, yamada_ref01, yamada_ref02,yamada_ref08, yamada}. By combining these techniques with scanning tunneling microscopy (STM), one can obtain complete information on the three-dimensional shape and electronic state of the molecule. Attempts have been made to obtain structural information from PMMs, specifically, the azimuthal alignment of the molecule \cite{yamada_ref03}, the orientation of the molecule in the multilayer \cite{yamada}, and the adsorption position of molecules based on multiple scattering \cite{niki_vac}. have been reported. It has been reported that there are still challenges in analyzing complex molecular thin film surfaces \cite{naturecom}. The combination of wavenumber-resolved photoelectron spectroscopy and a pulsed light source provides a PMM with femtosecond time resolution. 
\par
Structural measurement methods such as STM and X-ray standing wave (XSW) are difficult to combine to conduct a detailed electronic structure analysis because of the large-sized equipment and because it is necessary to simultaneously clarify the electronic state and structural dynamics from only one PMM to elucidate the photoexcitation process of electrons in molecules \cite{naturecom}. In this study, we developed a new analysis method to simultaneously estimate molecular orbital structure and molecular adsorption structure from a PMM.
\par
According to \textit{Fermi's Golden Rule}, the intensity of PMM $I(\vec{k})$ is expressed as 
\begin{linenomath}
\begin{align}
	I({\vec{k}}) \sim
	|\braket{\psi_{\vec{k}} | \hat{\Delta} | \psi_i(\vec{r})} |^2 
	\delta(E_i + \hbar \omega - E_{f} - \epsilon_{\vec{k}}), 
	\label{eq:I(k)}
\end{align}
\end{linenomath}
where $\psi_{\vec{k}}$ is the final state of photoelectron with momentum vector $\vec{k}$ and kinetic energy $\epsilon_{\vec{k}}$. $\vec{r}=(x, y, z)$ is an arbitrary point on the molecular orbital. $\hat{\Delta}$ is the electron-photon interaction operator, the $\hbar \omega$ is the energy of incident monochromatic photon, and $E_{f}$ is the energy state of the final state. When observing photoemission from molecular orbitals, the initial state $\psi_i$ is specific with $E_i$. Assuming that the orbitals of an $N$-atom molecule at $\vec{r}$ are represented by the \textit{linear combination of an atomic orbital (LCAO)} $\chi_\mu$: 
\begin{linenomath}
\begin{align}
	\psi_i({\vec{r}}) = 
	\sum_{n=1}^N \sum_\mu c_{i,n,\mu} \chi_\mu(\vec{r}-\vec{R}_n), 
	\label{eq:psi_i}
\end{align}
\end{linenomath}
where $c_{i,n,\mu}$ are the molecular orbital coefficients. $i$ and $\mu$ are orbital numbers of the molecular and atom, respectively. The $\mu$th atomic orbital $\chi_\mu$ is represented by some basis (plane wave basis, Slater basis, and Gaussian basis) representing the atomic orbital approximately. 
$\vec{R}_n=(X_n, Y_n, Z_n)$ is the position vector of the $n$th atom. 
\par
On the other hand, it has been proposed to use a plane wave \cite{Science, yamada_ref08} or a scattered wave \cite{MS01, MS02} as the final state $\psi_{\vec{k}}$. If a plane wave is used for the final state, Eq. (\ref{eq:I(k)}) can be expressed as 
\begin{linenomath}
\begin{align}
	I({\vec{k}}) \sim \Delta^2|\braket{e^{\I {\vec{k}}\cdot{\vec{r}}} | \psi_i({\vec{r}})}|^2, 
	\label{eq:I(k)2}
\end{align}
\end{linenomath}
where $\I=\sqrt{-1}$ denotes the imaginary unit. 
\par
Based on the plane wave approximation \cite{Science, yamada_ref08}, we can extract the electron-photon interaction term $\Delta$ from the absolute value. The intensity \textit{I} is then expressed as a square of the absolute value of the Fourier transform (FT) of the initial state $\psi_i$ represented by the intensity and phase. The initial state that ejects a photoelectron can be obtained through the inverse FT of a PMM; this method is called \textit{photoelectron orbital tomography} \cite{Science, yamada_ref01, yamada_ref02,yamada_ref08, yamada}. The POT from an experimental PMM has been performed with a trial phase as the initial value based on the Gerchberg-Saxton (GS) algorithm \cite{GS}. Generally, in experiments, the intensity \textit{I} lost phase information of an orbital because it is a square of the absolute value. Therefore, POT can be thought of as a phase retrieval problem.  
\par
In this paper, we propose a new method of POT using PMMs. Although a detailed analysis of molecular orbital shapes and energies from PMMs has been reported in previous POT-related studies \cite{Science, yamada_ref01, yamada_ref02,yamada_ref08, yamada, naturecom}, our methodology has three advantages. First, we established a method to estimate molecular orbitals with three-dimensional phases from a single PMM. Second, we show that our method is robust to noise, which is unavoidable in experiments. Finally, we derived a new concept for determining the adsorption-induced three-dimensional shape of a molecule from PMMs. We propose two methods: “structure discrimination," which actively uses expected structural information, and “structure estimation," which does not use any structural information.

\section{Method}
\par
In this paper, we propose to use the PhaseLift method, which was proposed by Cand\`es \textit{et al.} \cite{PLadd, Candes2013a}, to solve the problem to find sparse molecular orbital coefficients.

\subsection{Preliminaries}
\par
For a given length of the wavenumber vector $k=|\vec{k}|=|(k_x, k_y, k_z)|$, a PMM with specific kinetic energy determined by $k$ is given as a function of the wavenumber $\vec{k}= (k_x, k_y, (k^2-k_x^2-k_y^2)^{1/2})$. Let $\mathcal{K}=\{\vec{k}_1, \cdots, \vec{k}_M\}$ be a set of the wavenumber vectors for which a PMM is measured. 
\par
To make the number of the estimated molecular orbital coefficients finite, we restricted ourselves to the range of the orbital numbers of each atom in (\ref{eq:psi_i}) depending on the molecular orbital that we were interested in. Let $\mathcal{M}_n$ be a set of the possible orbital numbers for the $n$th atom, e.g., $\{1s\}$ for hydrogen and $\{1s, 2s, 2p_x, 2p_y, 2p_z\}$ for oxygen in the case where we select some atomic orbitals which combine to form highly occupied molecular orbitals. we also include atomic orbitals with deeper binding energies. Applying this restriction of the range of the orbital numbers of the $\mu$th atom as $\mathcal{M}_n$, (\ref{eq:psi_i}) becomes 
\begin{linenomath}
\begin{align}
	\psi_i({\vec{r}}) 
	= \sum_{n=1}^N \sum_{\mu \in \mathcal{M}_n} c_{i,n,\mu} \chi_\mu(\vec{r}-\vec{R}_n). 
	\label{eq:restricted-psi_i}
\end{align}
\end{linenomath}
\par
Let $z_m = I(\vec{k}_m)$ be a photoelectron intensity of the $m$th wavenumber vector $\vec{k}_m \in \mathcal{K}$. Substituting (\ref{eq:restricted-psi_i}) into (\ref{eq:I(k)2}), $z_m$ is given by 
\begin{linenomath}
\begin{align}
	z_m 
	& = 
	\Delta^2 \biggl| \sum_{n=1}^N \sum_{\mu \in \mathcal{M}_n} c_{i, n, \mu} 
	\int_{\mathbb{R}} d\vec{r} e^{-\I \vec{k}_m \cdot \vec{r}} 
	\chi_\mu(\vec{r}-\vec{R}_n) \biggr|^2 \\
	& = 
	\Delta^2 \biggl| \sum_{n=1}^N \sum_{\mu \in \mathcal{M}_n} a_{m, n,\mu} c_{i, n, \mu} 
	\biggr|^2, 
	\label{eq:zm}
\end{align}
\end{linenomath}
where we put $a_{m, n,\mu} = \int_{\mathbb{R}} d\vec{r} e^{-\I \vec{k}_m \cdot \vec{r}} \chi_\mu(\vec{r}-\vec{R}_n)$, which is a value of the Fourier transformation of the atomic orbital at the wavenumber vector $\vec{k}_m$. PMM can be then represented as an $M$-dimensional column vector 
\begin{linenomath}
\begin{align}
	\vec{z} = (z_1, \cdots, z_M)^\top, 
\end{align}
\end{linenomath}
where $\vec{z}^\top$ denotes the transpose of $\vec{z}$. Let $L = \sum_{n=1}^N |\mathcal{M}_n|$ be the number of the molecular orbital coefficients $\{ c_{i, n, \mu} \}$, where $|\mathcal{M}_n|$ stands for the cardinarity of the set $\mathcal{M}_n$. 
\par
We next introduce a matrix $A \in \mathbb{C}^{M \times L}$ and a vector $\vec{c} \in \mathbb{R}^L$. We represented the entries in the set $\mathcal{M}_\mu$ as $\{\mu_1, \cdots, \mu_{|\mathcal{M}_n|}\}$. The matrix $A$ is defined by the following $M \times N$-block matrix: 
\begin{linenomath}
\begin{align}
	& A = \left(
	\begin{array}{c|c|c}
		\vec{a}_{1,1} & \quad\cdots\quad & \vec{a}_{1,N} \\
		\hline 
		\vdots & & \vdots \\
		\hline 
		\vec{a}_{M,1} & \quad\cdots\quad & \vec{a}_{M,N} \\
	\end{array}
	\right), 
	\label{eq:A}
\end{align}
\end{linenomath}
where the $(m,n)$th block of the matrix $A$ is an $|\mathcal{M}_n|$-dimensional row vector $\vec{a}_{m,n} = (a_{m, n,\mu_1}, \cdots, a_{m, n, \mu_{|\mathcal{M}_n|}})$. The matrix $A$ is referred to as the \textit{measurement matrix}. In a similar way, the vector $\vec{c}$ is defined by the following $N$-block column vector: 
\begin{linenomath}
\begin{align}
	\vec{c} = (\vec{c}_1^\top\; | \quad\cdots\quad | \;\vec{c}_N^\top)^\top, 
	\label{eq:c}
\end{align}
\end{linenomath}
where the $n$th block of the vector $\vec{c}$ is an $|\mathcal{M}_n|$- dimensional column vector $\vec{c}_n = (c_{i, n, \mu_1}, \cdots, c_{i, n, \mu_{|\mathcal{M}_n|}})^\top$. Let $\vec{a}_m^{\,*}$ be the $m$th row vector of the matrix $A$, where $\vec{a}_m^{\,*}$ denotes the conjugate transpose of $\vec{a}_m$. Using $A$ and $\vec{c}$, Eq. (\ref{eq:zm}) becomes $z_m = \Delta^2 | \vec{a}_m^{\,*} \vec{c}\, |^2$. PMM $\vec{z}$ can be then written as follows: 
\begin{linenomath}
\begin{align}
	\vec{z} = \Delta^2 | A \vec{c} \, |^2, 
	\label{eq:z}
\end{align}
\end{linenomath}
where the modulus is supposed to be entrywise applied for a vector, i.e., $|(x_1, \cdots, x_n)|=(|x_1|, \cdots, |x_n|)$. 
\par
The problem to obtain molecular orbital coefficients is equivalent to find $\vec{c}$ that satisfies (\ref{eq:z}), i.e., the following optimization problem: 
\begin{linenomath}
\begin{alignat}{3}
	& \text{find} && \vec{c} \notag\\
	& \text{subject~to~~~} && \vec{z} = \Delta^2 | A \vec{c} \, |^2, 
	\label{eq:original-phase-retrieval}
\end{alignat}
\end{linenomath}
for given $A$, $\vec{z}$ and $\Delta$. This kind of problems is called the \textit{phase retrieval} problem, and it is an NP-hard problem in general. It is therefore hard to solve (\ref{eq:original-phase-retrieval}) as it is.

\subsection{PhaseLift}
\par
We briefly summarize the idea of the PhaseLift method. In the PhaseLift algorithm, the problem (\ref{eq:original-phase-retrieval}) is approximately transformed into the semidefinite programming (SDP). First, the $L$-dimensional variable $\vec{c}\,$ is embedded in a higher dimensional space using the transformation $C=\vec{c}\, \vec{c}^{\,*}$. Since the eigenvalues of $C$ are 0 and 1, $C$ is a positive semidefinite matrix. Note that each entry in $\vec{c}\,$ is actually a real value in this problem. This transformation is called the \textit{lifting}. Since $C^*=C$ holds, the matrix $C$ is an $L \times L$-Hermitian matrix. Using this transformation, the photoelectron intensity of the $m$th wavenumber vector becomes linear in the matrix $C$ as 
\begin{linenomath}
\begin{align}
	z_m 
	&= \Delta ^2| \vec{a}_m^* \vec{c}\, |^2 \notag\\
	&= \Delta^2 (\vec{a}_m^* \vec{c}\,)^* \; \vec{a}_m^* \vec{c}\, \notag\\
	&= \Delta^2 \mathrm{tr}( \vec{c}^{\,*} \vec{a}_m \vec{a}_m^* \vec{c}\, ) \notag\\
	&= \Delta^2 \mathrm{tr}( \vec{a}_m \vec{a}_m^* \vec{c}\, \vec{c}^{\,*} ) \notag\\
	&= \Delta^2 \mathrm{tr}( A_m C ), 
\end{align}
\end{linenomath}
where $A_m$ is the rank-one matrix $\vec{a}_m \vec{a}_m^*$. Here, $\mathrm{tr}(X)$ stands for the trace of a square matrix $X$, which is the sum of the diagonal entries. The problem (\ref{eq:original-phase-retrieval}) is equivalent to 
\begin{linenomath}
\begin{alignat}{3}
	& \text{find} && C \notag\\
	& \text{subject~to~~~} && z_m = \Delta^2 \mathrm{tr}( A_m C) \;
	~\text{for all}~ m \notag\\
	& && C \succeq O \notag\\
	& && \mathrm{rank}(C) = 1, 
	\label{eq:phase-retrieval}
\end{alignat}
\end{linenomath}
where $C \succeq O$ means that $C$ is a positive semidefinite matrix. The problem (\ref{eq:phase-retrieval}) is equivalent to the following reformulation: 
\begin{linenomath}
\begin{alignat}{3}
	& \text{min} && \mathrm{rank}(C) \notag\\
	& \text{subject~to~~~} && z_m = \Delta^2 \mathrm{tr}( A_m C) \; 
	~\text{for all}~ m \notag\\
	& && C \succeq O. 
	\label{eq:rank-minimization}
\end{alignat}
\end{linenomath}
However, this rank minimization problem (\ref{eq:rank-minimization}) is NP-hard. Therefore, the trace is used as a convex surrogate. Then a convex relaxation of (\ref{eq:rank-minimization}) becomes 
\begin{linenomath}
\begin{alignat}{3}
	& \text{min} && \mathrm{tr}(C) \notag\\
	& \text{subject~to~~~} && z_m = \Delta^2 \mathrm{tr}( A_m C) \;
	~\text{for all}~ m \notag\\
	& && C \succeq O. 
	\label{eq:trace-minimization}
\end{alignat}
\end{linenomath}
\par
In this way, we have arrived at a \textit{semidefinite program} (SDP) that is equivalent to a problem to obtain a molecular orbital using the lifting technique. It has been proved that, under certain conditions, if the trace minimization program (\ref{eq:trace-minimization}) returns a rank-one solution, then this solution is exact \cite{Candes2013a}. Even when such conditions do not hold, the trace minimization program returns a low-rank solution, and we can approximately obtain a rank-one solution as $\vec{c} \approx \sqrt{\lambda_1} \vec{u}_1$ using the maximum eigenvalue $\lambda_1$ and its eigenvector $\vec{u}_1$ by applying the eigenvalue decomposition $C = \sum_{\ell=1}^L \lambda_\ell \vec{u}_\ell\vec{u}_\ell^*$, where $\lambda_\ell$ and $\vec{u}_\ell$ are the eigenvalues and the eigenvectors of $C$ with $\lambda_1 \ge \cdots \ge \lambda_L \ge 0$.

\subsection{Robust sparse PhaseLift for noisy measurements}
\par
In many cases, PMMs contain the measurement noise. One can also find a sparse solution under noisy measurements. PhaseLift can be naturally reformulate for noisy measurement by relaxation of the condition and introducing the $\ell_1$ norm regularization term as follows. 
\begin{linenomath}
\begin{alignat}{3}
	& \text{min} && \lambda_{tr} \mathrm{tr}(C) 
	+ \lambda_{norm} \| C \|_1 \notag \\
    & && + \sum_{m=1}^M | z_m - \Delta^2 \mathrm{tr}( A_m C) | \notag\\
	& \text{subject~to~~~} && C \succeq O, 
	\label{eq:robust-sparse-phaselift}
\end{alignat}
\end{linenomath}
where $\lambda_{tr}$ and $\lambda_{norm}$ are parameters to control the tolerance of the measurement noise \cite{Hand2017} and $\| C \|_1$ stands for the $\ell_1$ norm for a matrix $C=(C_{\ell \ell'})$, i.e., $\| C \|_1 = \sum_{\ell=1}^L \sum_{\ell'=1}^L |C_{\ell \ell'}|$. In our settings, the system can be underdetermined or overdetermined, depending on how the measurement matrix $A$ is determined. This robust sparse PhaseLift (\ref{eq:robust-sparse-phaselift}) is valid in both cases.

\section{Numerical experiments}
\par
It is shown that our method effectively works by numerical experiments. Firstly, we show that it can find the molecule orbital from the PMM that is theoretically generated based on the plane wave approximation and Gaussian basis, which is  
\begin{linenomath}
\begin{align}
	\chi_\mu(\vec{r}) =& \sum_{p=1}^P d_{\mu, p} g_{{\mu,} p}({\vec{r}}), \notag \\
	g_{{\mu,} p}(\vec{r}) =& {\gamma_{\mu, p}} 
	x^\ell y^m z^n e^{-\alpha_{\mu, p} \|\vec{r}\|^2}. 
\end{align}
\end{linenomath}
$d_{\mu, p}$ and $\alpha_{\mu, p}$ are atomic orbital coefficient and Gaussian coefficient. $p$ denotes the basis function number, and each coefficient is a constant determined for each entry. $x^\ell y^m z^n$ determines the orbital shape of the atom by the difference in ``$l, m, n$''. The coefficients $\gamma_{\mu, p}$ is a normalization constant to satisfy $\int_{\mathbb{R}^3} \{g_{\mu, p}(\vec{r})\}^2 d\vec{r} = 1$, i.e., $\gamma_{\mu, p} = [2^{2(\ell+m+n)+3/2} \alpha^{\ell+m+n+3/2} / \{(2\ell -1)!!(2m -1)!!(2n -1)!!\pi^{3/2}\}]^{1/2}$. For instance, $P=3$ in the case where STO-3G is used as the Gaussian basis. We set $\lambda_{tr}=1$ and $\lambda_{norm}=0.01$ for all numerical experiments. 
\par
It should be noted that the computational cost can be reduced using symmetric property of a molecule. If molecule's structure has a certain symmetry such as F4-TCNQ, there exist some sets of atoms that can be assumed to take same molecular orbital coefficients. In such a case, since the PMM is given via linear combination of bases, the number of molecular orbital coefficients to be estimated can be reduced. For simplicity, let $A=(\vec{\phi}_1, \cdots, \vec{\phi}_L) \in \mathbb{C}^{M \times L}$ and $\vec{c}=(c_1, \cdots, c_L)^\top \in \mathbb{C}^L$ be a measurement matrix and a molecular orbital coefficients, respectively. Here, $\vec{\phi}_\ell$ is the $\ell$th column vector of $A$. For example, suppose that the molecular coefficients $c_1$ and $c_2$ can be expected to have the same value, i.e., $c_1=c_2$. Then, $A \vec{c} = c_1 \vec{\phi}_1 + c_2 \vec{\phi}_2 + c_3 \vec{\phi}_3 + c_L \vec{\phi}_L = c_1 (\vec{\phi}_1 + \vec{\phi}_2) + c_3 \vec{\phi}_3 + c_L \vec{\phi}_L = A' \vec{c}\,'$ holds, where $A'=(\vec{\phi}_1+\vec{\phi}_2, \vec{\phi}_3, \cdots, \vec{\phi}_L) \in \mathbb{C}^{M \times (L-1)}$ and $\vec{c}\,'= (c_1, c_3, \cdots, c_L) \in \mathbb{C}^{L-1}$. As a result, the dimension of the vector to be estimated $\vec{c}$ was reduced by one. Similarly, if there are $k$ orbital molecular coefficients are expected to be the same, the dimension of $\vec{c}$ can be reduced by $k-1$. Furthermore, this process can be repeated for all sets of the molecular orbital coefficients that take the same value. This operation reduces the computational cost.

\subsection{Phase retrieval from a theoretical PMM using sparse PhaseLift}
\par
First, we selected sample molecules which determined the elements, the number of atoms, and coordinate vector $\vec{R} = (R_x, R_y, R_z)$ of each atom accordingly. Then, to calculate theoretical PMMs based on eq. (\ref{eq:z}), we first chose the Gaussian basis for the atomic orbitals $\chi_\mu$. Next, the molecular orbital coefficients $\vec{c}_0$ and the coefficients $(d_{\mu p}, \alpha_{\mu p}, l, m, n)$ associated with the atomic orbitals $\chi_\mu$ were obtained using the PySCF \cite{pyscf_01, pyscf_02, pyscf_03}. Based on the definition of the matrix $A$, the $M \times N$ block matrix $A$ was constructed using the coefficients obtained above. From all of the procedure, the theoretical PMMs were successfully obtained. 
\par
We consider a PMM $\vec{z}$ that is generated by a given molecular orbital coefficients $\vec{c}_0$, i.e., 
\begin{linenomath}
\begin{align}
	\vec{z} = \Delta^2 | A \vec{c}_0 \, |^2, 
	\label{eq:theoretical-PMM}
\end{align}
\end{linenomath}
which is referred to as the \textit{theoretical PMM}. The error between the true molecular orbital $\psi_i({\vec{r}})$ determined by $\vec{c}\,_0$ and the estimated molecular orbital $\tilde{\psi}_i(\vec{r})$ is defined by 
\begin{linenomath}
\begin{align}
	\epsilon_{MO}(\psi_i, \tilde{\psi}_i) = 1 - 
	\frac{\braket{\psi_i | \tilde{\psi}_i}}
	{(\braket{\psi_i | \psi_i}
	\braket{\tilde{\psi}_i | \tilde{\psi}_i})^{1/2}}, 
\end{align}
\end{linenomath}
where the second term in the right-hand side is the direction cosine between the true and the estimated molecular orbitals. The error between the true molecular orbital and the estimated molecular orbital is referred to as the \textit{MO-error}. We also measure the distance between two PMMs $\vec{z}$ and $\vec{z}\,'$ by the total variation distance (TVD) of their normalized vectors: 
\begin{linenomath}
\begin{align}
	d_{TV}(\vec{z}, \vec{z}\,') = 
	\biggl\| \frac{\vec{z}}{\|\vec{z}\|_1} - 
	\frac{\vec{z}\,'}{\|\vec{z}\,'\|_1} \biggr\|_1. 
	\label{eq:TV}
\end{align}
\end{linenomath}
Although the definition of the total variation distance for two probability vectors sometimes has the factor $1/2$ to constrain it within the interval of $[0,1]$, it is often omitted as in (\ref{eq:TV}). 
For all theoretical PMMs in this paper, we calculated the photoelectron energy as 55 eV.
\par
Figures 1, 2, and 3 show the valence molecular orbitals calculated using PySCF, the theoretical PMMs based on the plane wave approximation, the estimated PMMs by our method, and the estimated molecular orbitals. To verify the accuracy of our method, we first performed calculations for F4-tetracyanoquinodimethane (F4-TCNQ) (Figs. 1(a1)-(a6)). Estimating charge transfer from the metal to each molecule has been discussed even for smaller molecules such as benzene \cite{F4TCNQ_ref8,F4TCNQ_ref9,F4TCNQ_ref10}. We chose F4-TCNQ because of the adsorption-induced geometric distortion and electron donation and back donation, which directly affect the occupations of molecular orbitals and the metal work function \cite{F4TCNQ}. Figure 1(a3) shows a theoretical PMM with photoemission intensities calculated from the $2p$ orbital based on the plane wave approximation. We can confirm that photoelectrons are emitted in the direction of the molecular orbital (Fig. 1(a1)) using PySCF. Figures 1(a4)-(a6) show the estimated PMM and estimated molecular orbital using our method, respectively. These results are in good agreement with the theoretical PMM (Fig. 1(a3)) and the molecular orbital (Figs. 1(a1) and (a2)). The respective MO-error is 0.0004. 
\par
A similar trend can be seen for $\mathrm{H_2O}$ (Figs. 2(a1)-(a4)), $\mathrm{H_2SO_4}$ (Figs. 2(b1)-(b4)), acetamide (Figs. 2(c1)-(c4)), acetic acid (Figs. 2(d1)-(d4)), one of the degenerate orbitals of benzene (Figs. 2(e1)-(e4)), and anirin (Figs. 2(f1)-(f4)). The respective MO-errors were 0.0000 ($\mathrm{H_2O}$), 0.0144 ($\mathrm{H_2SO_4}$), 0.0264 (acetamide), 0.0417 (acetic acid), 0.0000 (benzene), and 0.0068 (anirin). Molecules have a steric structure in which molecular bonds extend beyond the paper plane. Comparing theoretical PMMs according to molecular orbital PySCF, photoelectron emission is observed in the direction of the extended molecular orbital in all calculations. As aniline and benzene have similar molecular orbitals, the theoretical PMM for benzene is similar to that for aniline. These PMMs are divided into two because their molecular orbitals bisect the molecule. F4-TCNQ shows a similar trend. 
\par
These results exhibit MO-error values lower than 0.04, indicating that our method can estimate the three-dimensional molecular orbitals and their phases from a single theoretical PMM. On the other hand, as shown in Figures 3(a1)-(a4), the MO-error for $\mathrm{N_2}$ (0.0004) is larger than that for $\mathrm{H_2O}$, indicating that the symmetry in molecular height makes estimation difficult. The distance between two PMMs between each theoretical PMM and the estimated PMM are 0.0004 (F4-TCNQ), 0.0000 ($\mathrm{H_2O}$), 0.0004 ($\mathrm{H_2SO_4}$), 0.0003 (acetamide), 0.0006 (acetic acid), 0.0002 (benzene), 0.0008 (anirin), and 0.0004 ($\mathrm{N_2}$). The estimated molecular orbitals are in good agreement with those calculated using PySCF. These results suggest that our method succeeded in estimating molecular orbitals and their phases. This approach is suitable for estimating molecular orbitals from theoretical or experimental PMMs.
\par
\begin{figure*}[t]
    \centering
	\begin{minipage}{0.15\linewidth}
		\centering
		\includegraphics[width=1\linewidth]{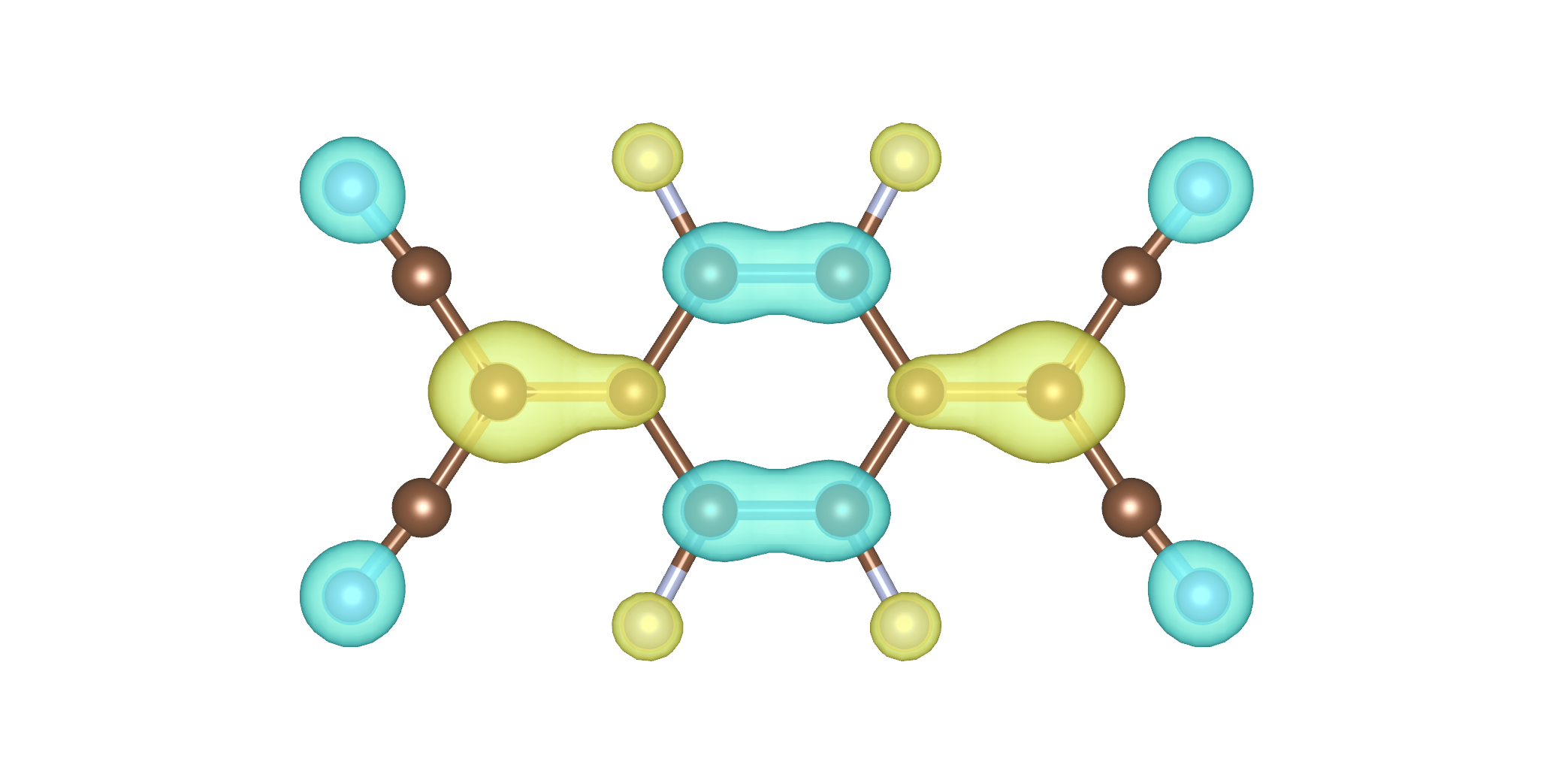}\\[-0.8em]
		\footnotesize (a1)
	\end{minipage}
    \begin{minipage}{0.15\linewidth}
		\centering
		\includegraphics[width=1\linewidth]{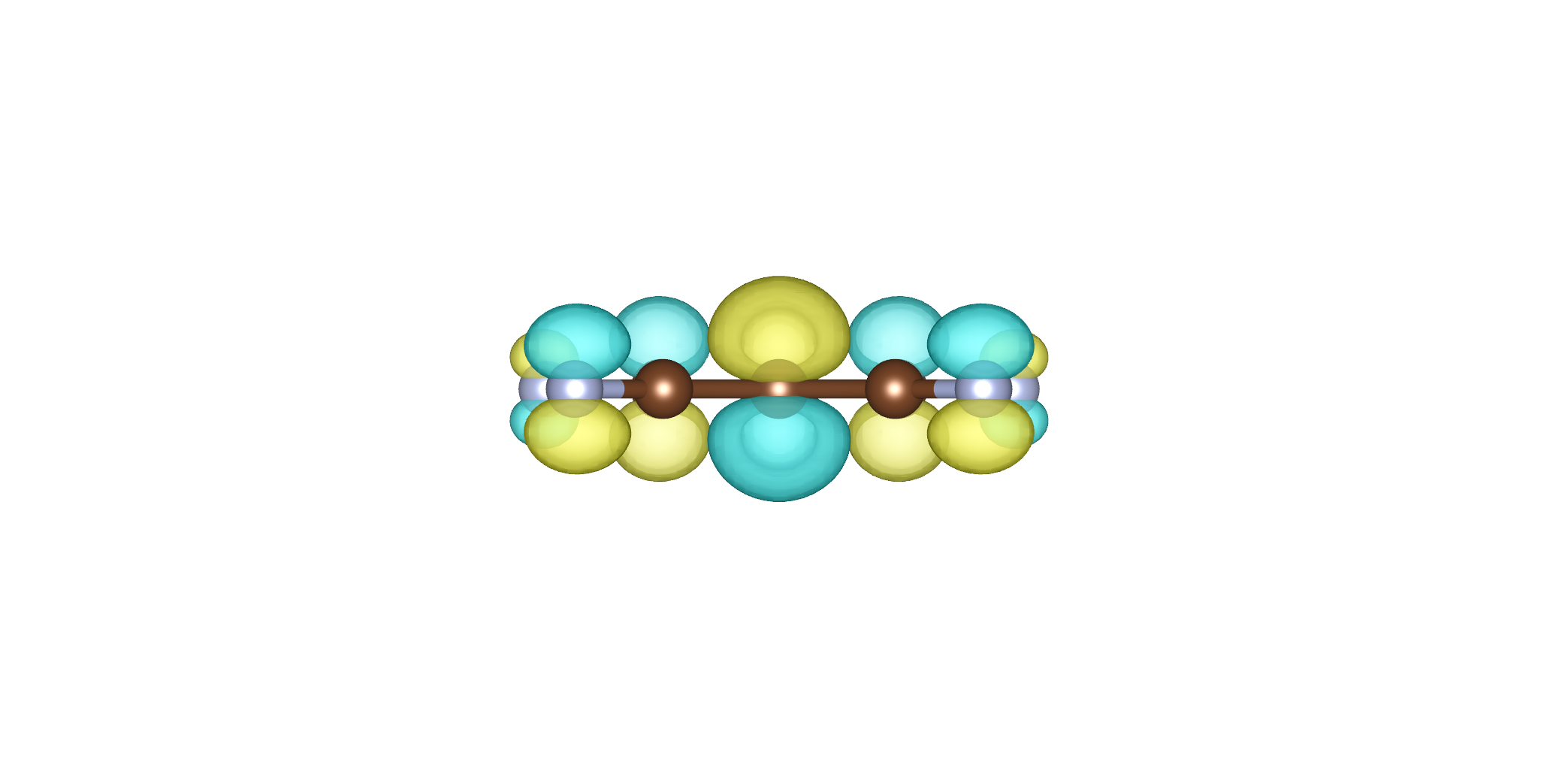}\\[-0.8em]
		\footnotesize (a2)
	\end{minipage}
	\begin{minipage}{0.15\linewidth}
		\centering
		\includegraphics[width=0.8\linewidth]{./fig.theoretical_PAD_F4TCNQ}\\[-0.8em]
		\footnotesize (a3)
	\end{minipage}
	\begin{minipage}{0.15\linewidth}
		\centering
		\includegraphics[width=0.8\linewidth]{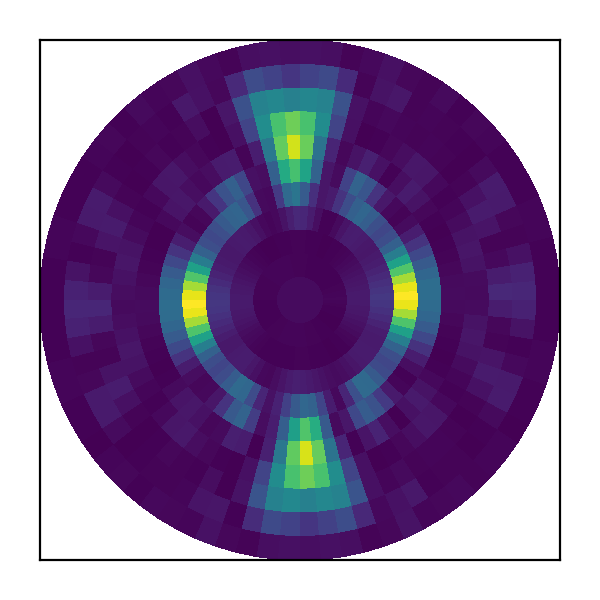}\\[-0.8em]
		\footnotesize (a4)
	\end{minipage}
	\begin{minipage}{0.15\linewidth}
		\centering
		\includegraphics[width=1\linewidth]{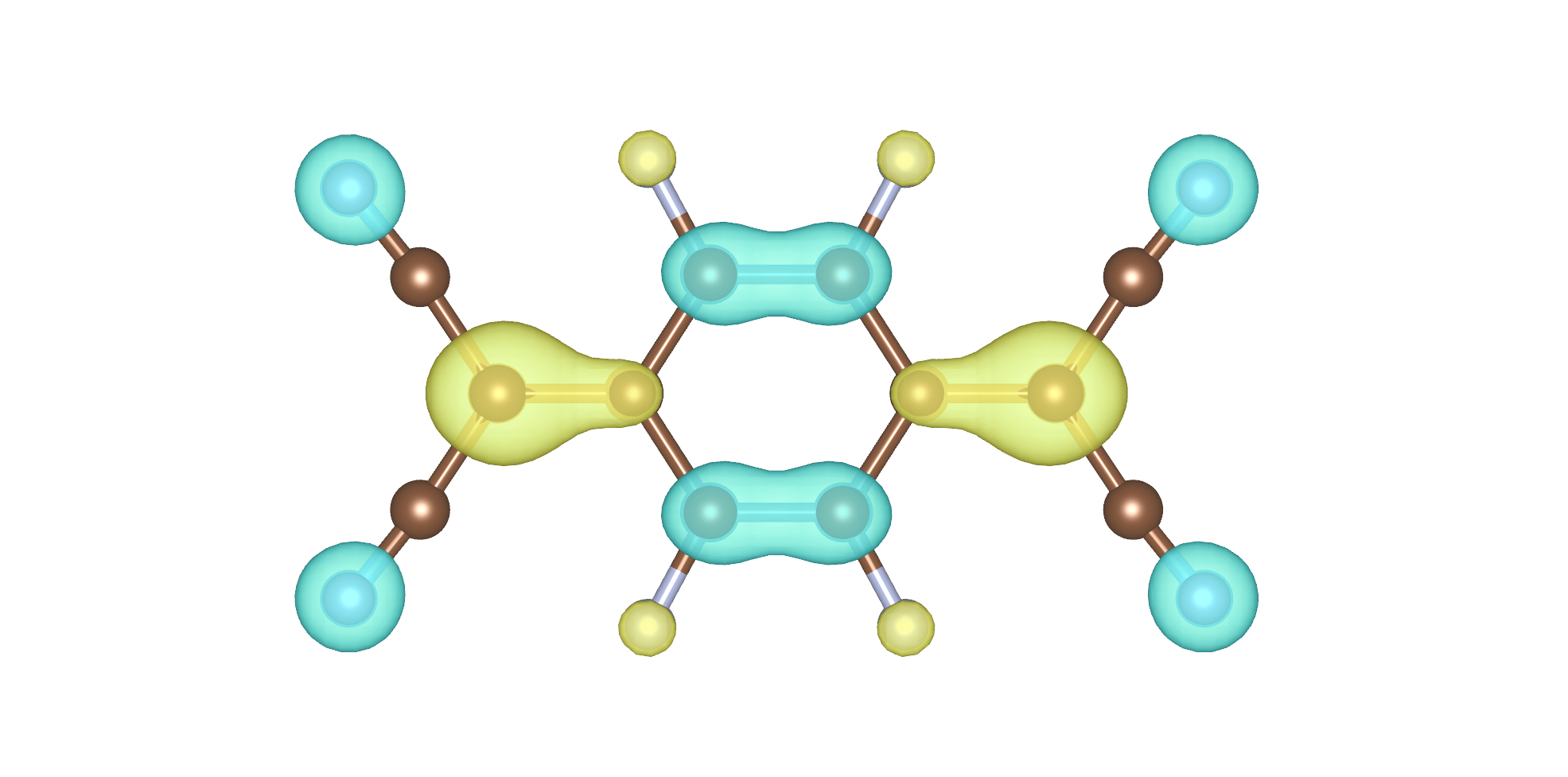}\\[-0.8em]
		\footnotesize (a5)
	\end{minipage}
    \begin{minipage}{0.15\linewidth}
		\centering
		\includegraphics[width=1\linewidth]{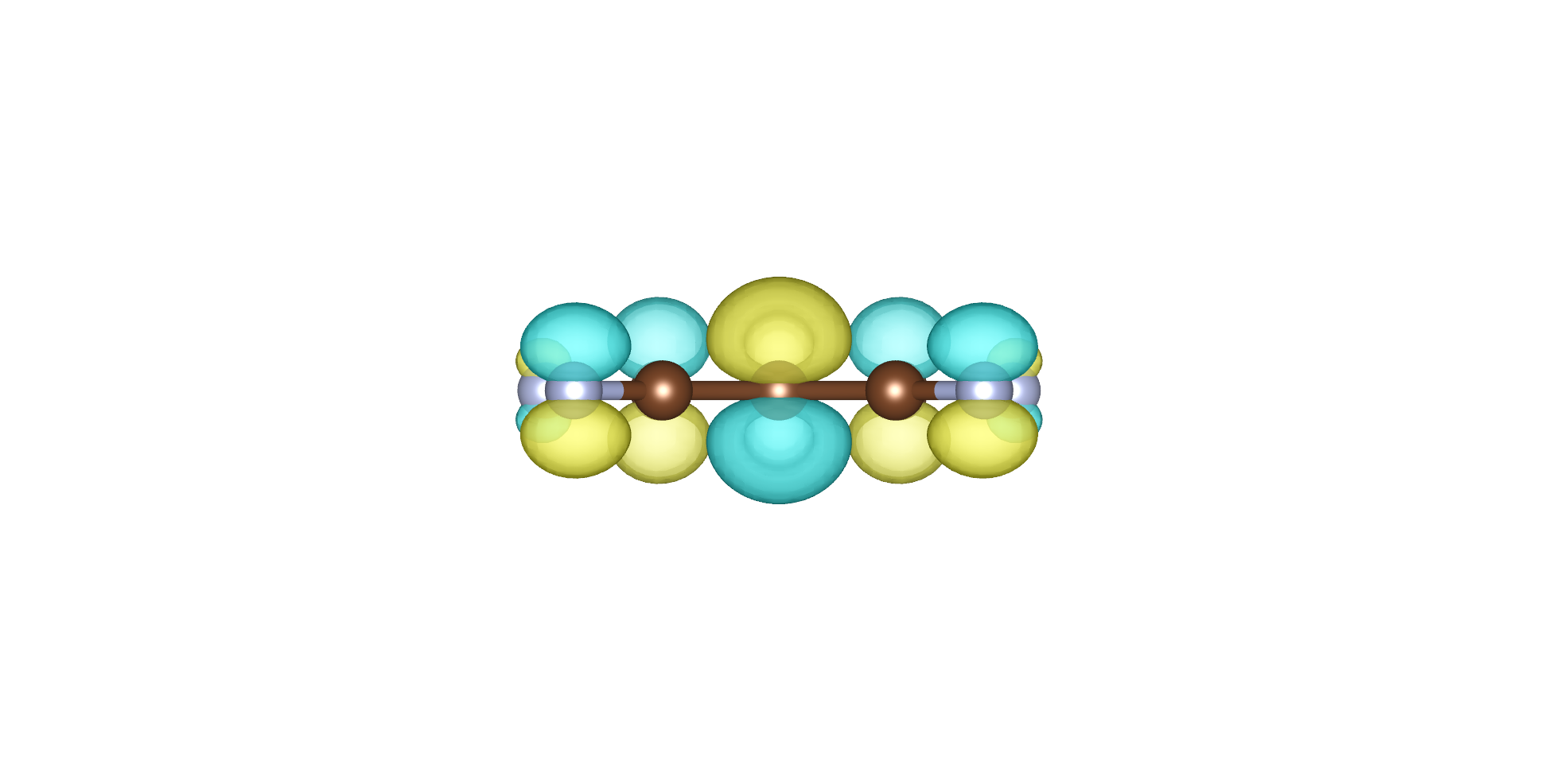}\\[-0.8em]
		\footnotesize (a6)
	\end{minipage}
	\\
	\caption{PMMs of F4TCNQ and it's molecule orbitals. A PMM in $k_x k_y$-plane is shown where $k_x$ along the horizontal axis and $k_y$ along the longitudinal axis range from -4 to 4 [1/\AA]. The center position of the PMM corresponds to $(k_x, k_y) = (0,0)$. The maximum (minimum) intensity is represented by yellow (blue). 		 
	}
	\label{fig:simple-molecules_cont'd}
\end{figure*}
\par
\begin{figure*}[t]
	\centering
	\begin{minipage}{0.15\linewidth}
		\centering
		\includegraphics[width=1\linewidth]{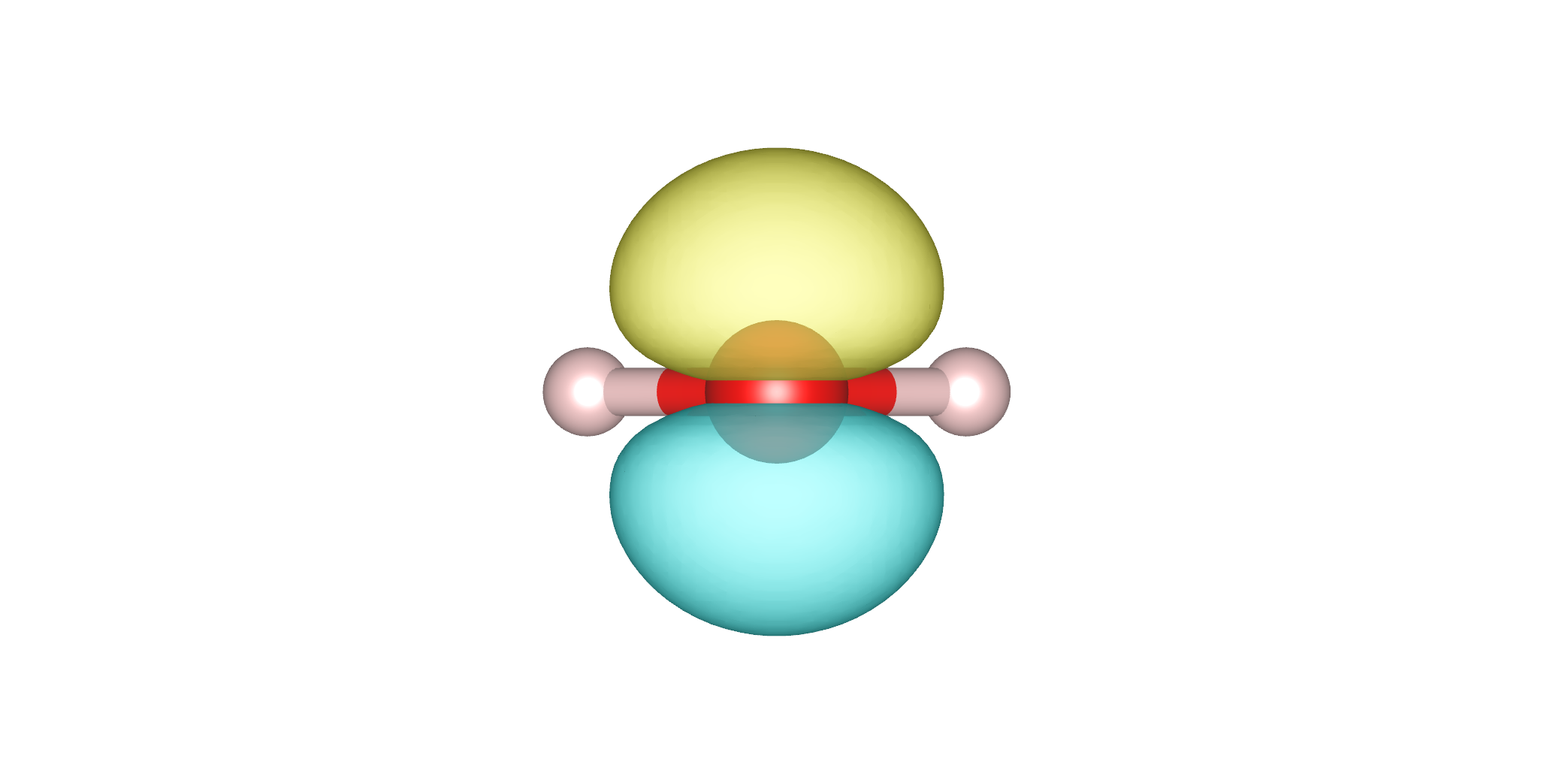}\\[-0.8em]
		\footnotesize (a1)
	\end{minipage}
        \begin{minipage}{0.15\linewidth}
		\centering
		\includegraphics[width=1\linewidth]{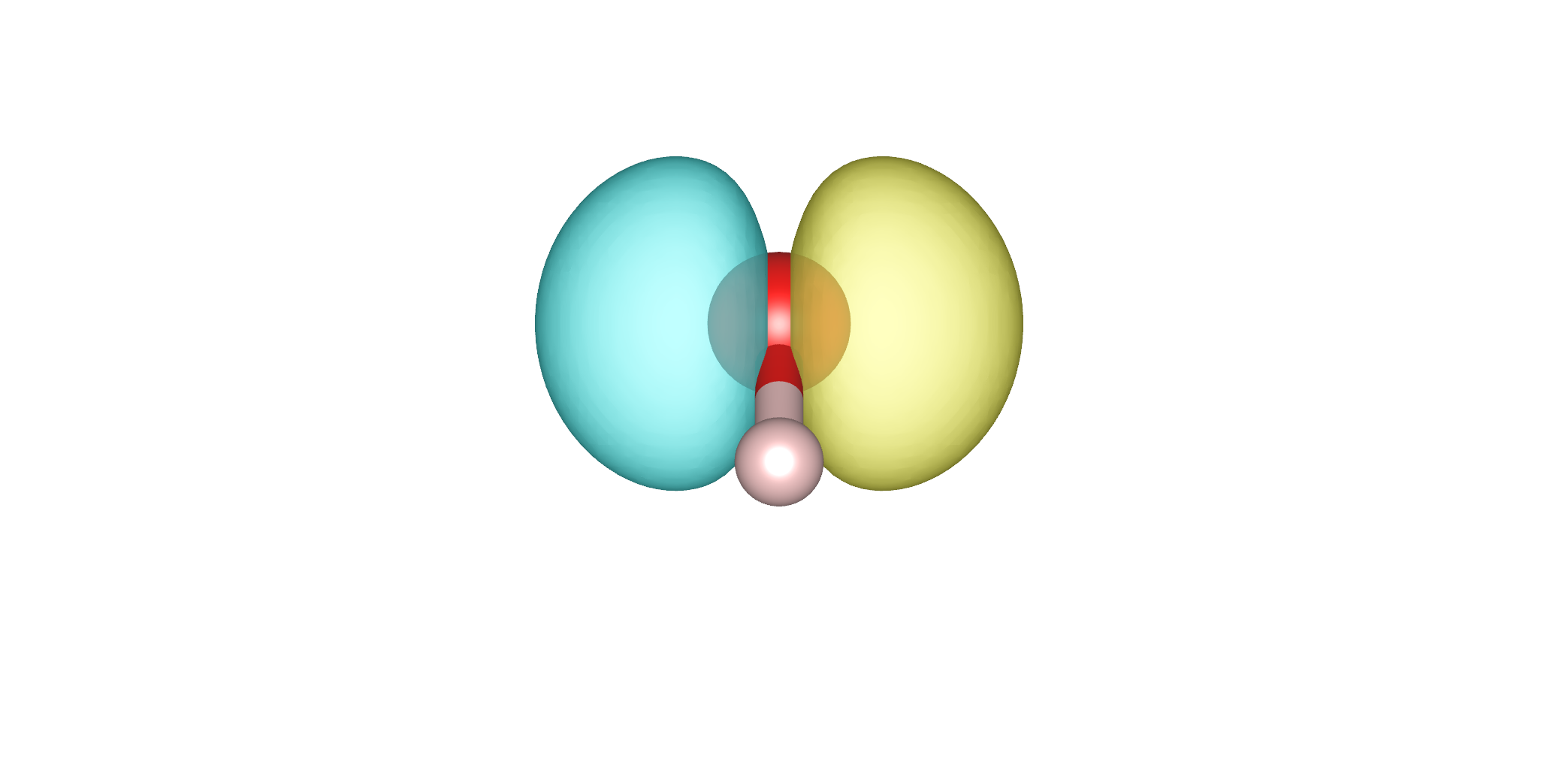}\\[-0.8em]
		\footnotesize (a2)
	\end{minipage}
	\begin{minipage}{0.15\linewidth}
		\centering
		\includegraphics[width=0.8\linewidth]{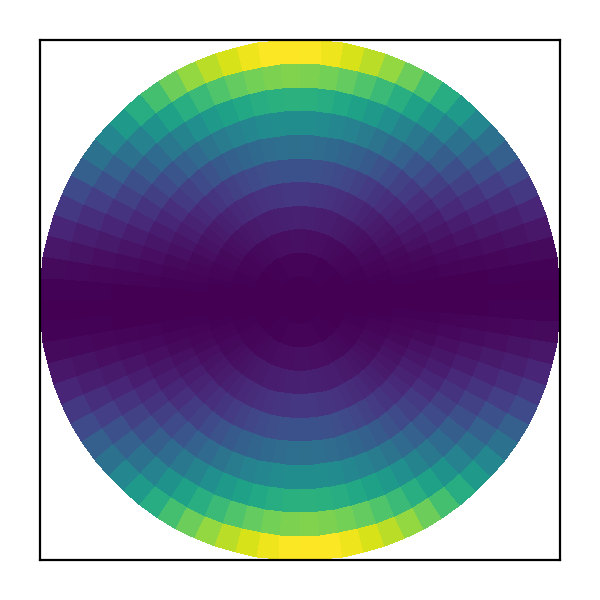}\\[-0.8em]
		\footnotesize (a3)
	\end{minipage}
	\begin{minipage}{0.15\linewidth}
		\centering
		\includegraphics[width=0.8\linewidth]{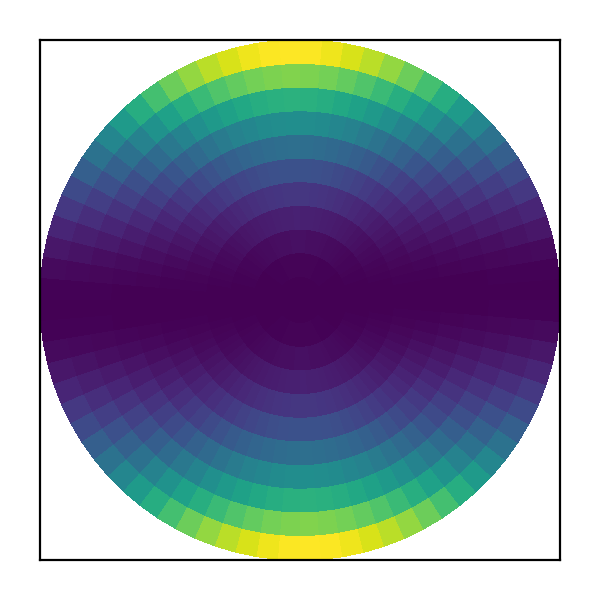}\\[-0.8em]
		\footnotesize (a4)
	\end{minipage}
	\begin{minipage}{0.15\linewidth}
		\centering
		\includegraphics[width=1\linewidth]{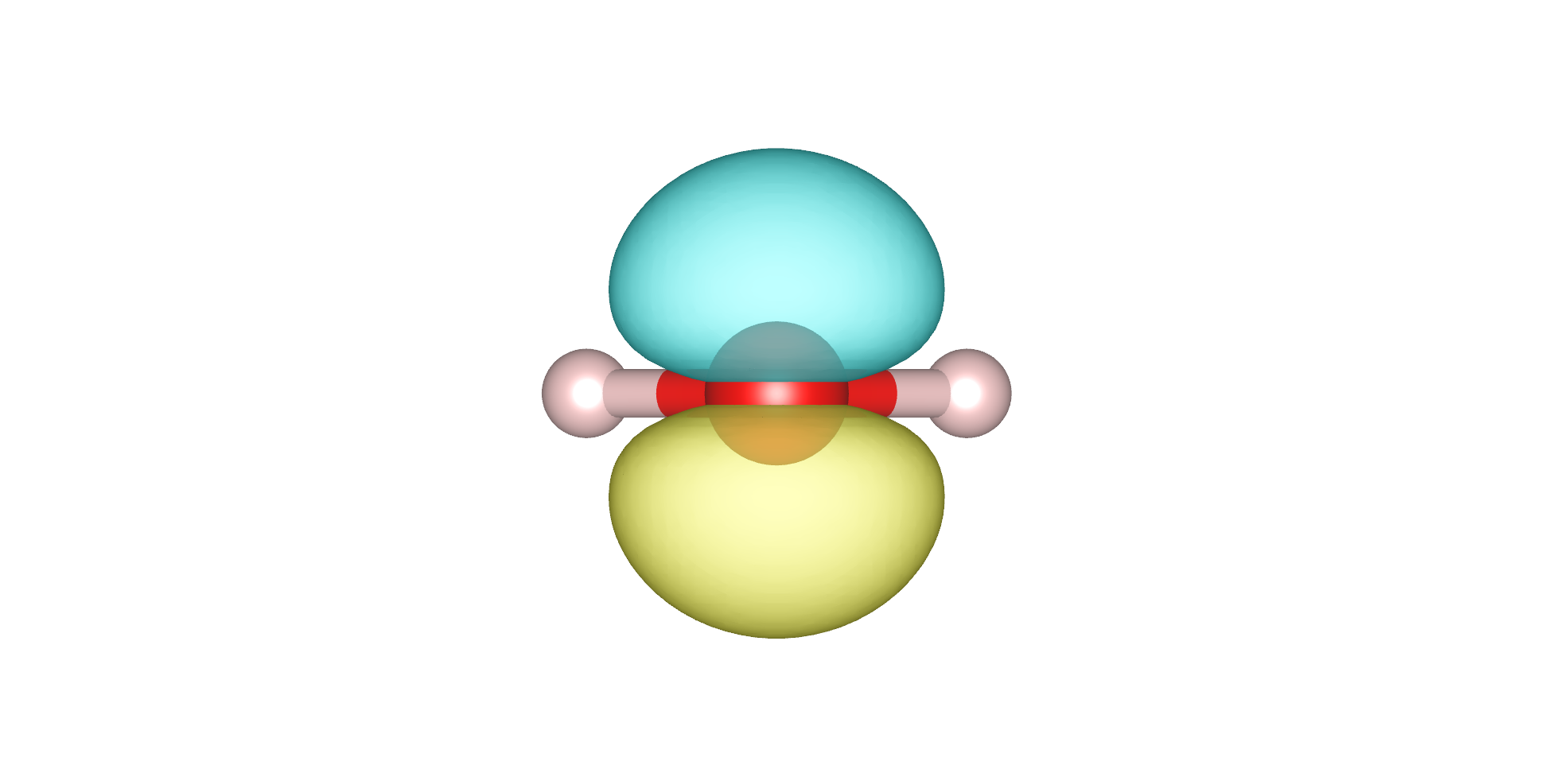}\\[-0.8em]
		\footnotesize (a5)
	\end{minipage}
        \begin{minipage}{0.15\linewidth}
		\centering
		\includegraphics[width=1\linewidth]{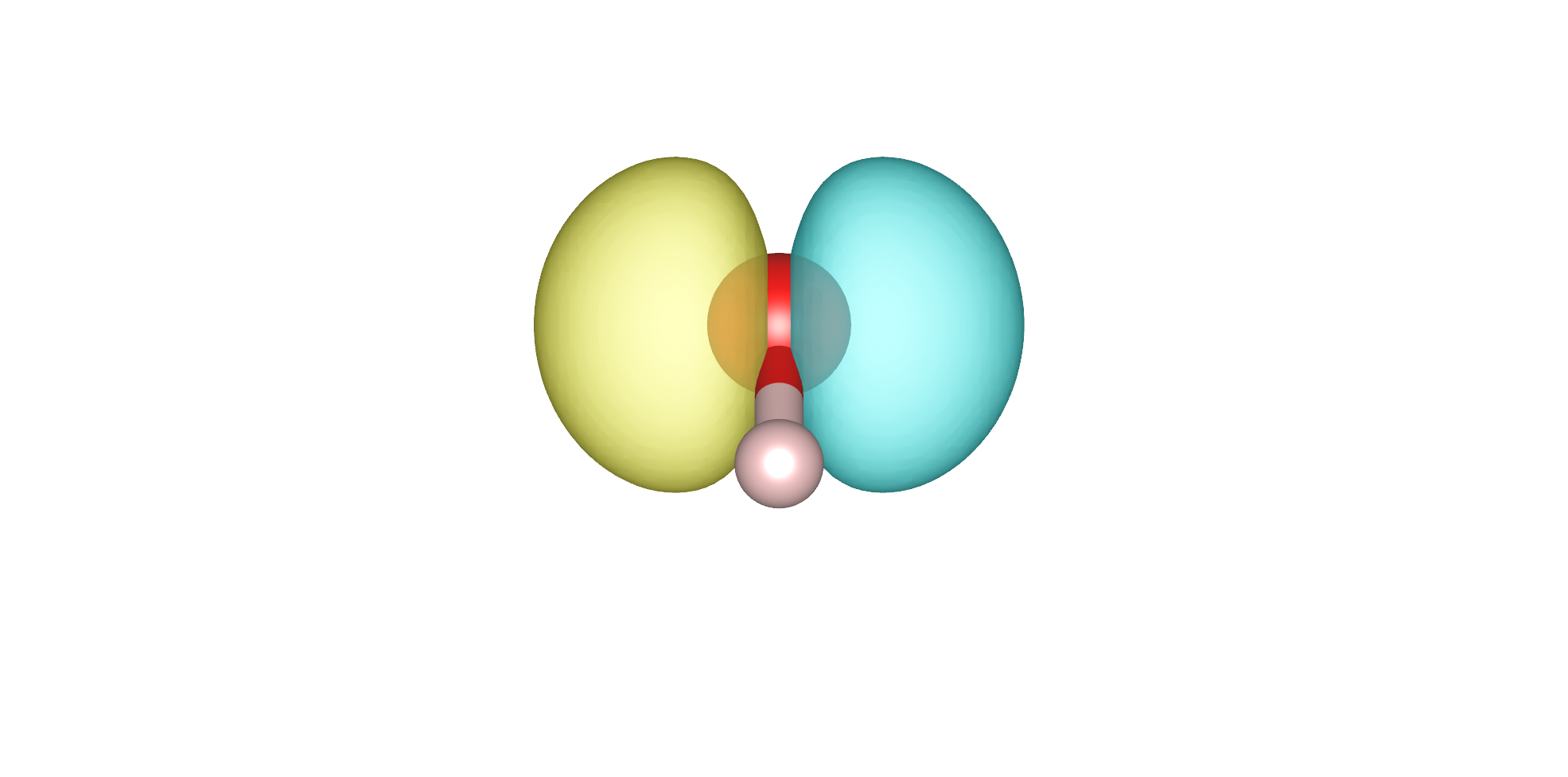}\\[-0.8em]
		\footnotesize (a6)
	\end{minipage}
	\\
	\begin{minipage}{0.15\linewidth}
		\centering
		\includegraphics[width=1\linewidth]{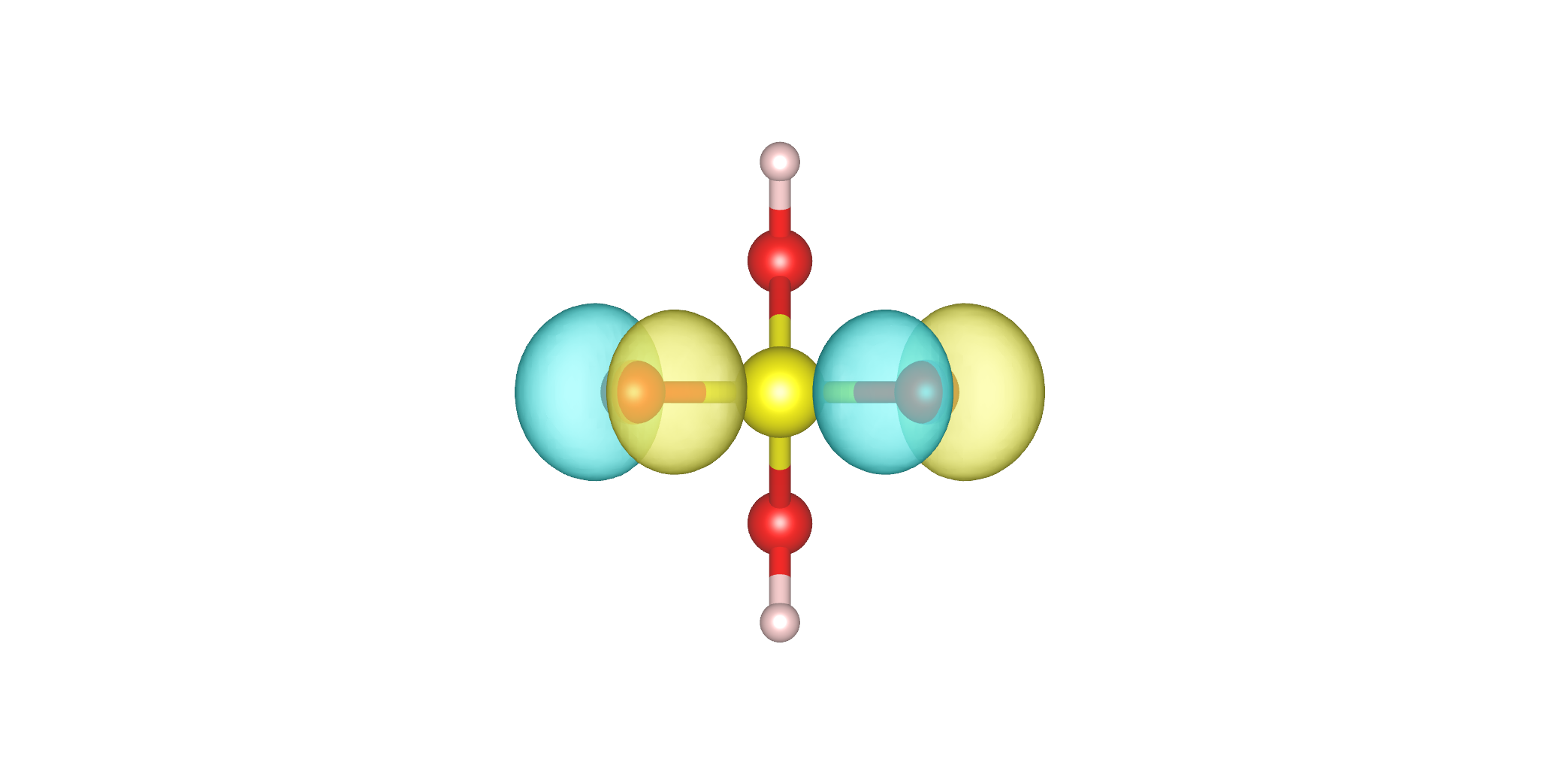}\\[-0.8em]
		\footnotesize (b1)
	\end{minipage}
        \begin{minipage}{0.15\linewidth}
		\centering
		\includegraphics[width=1\linewidth]{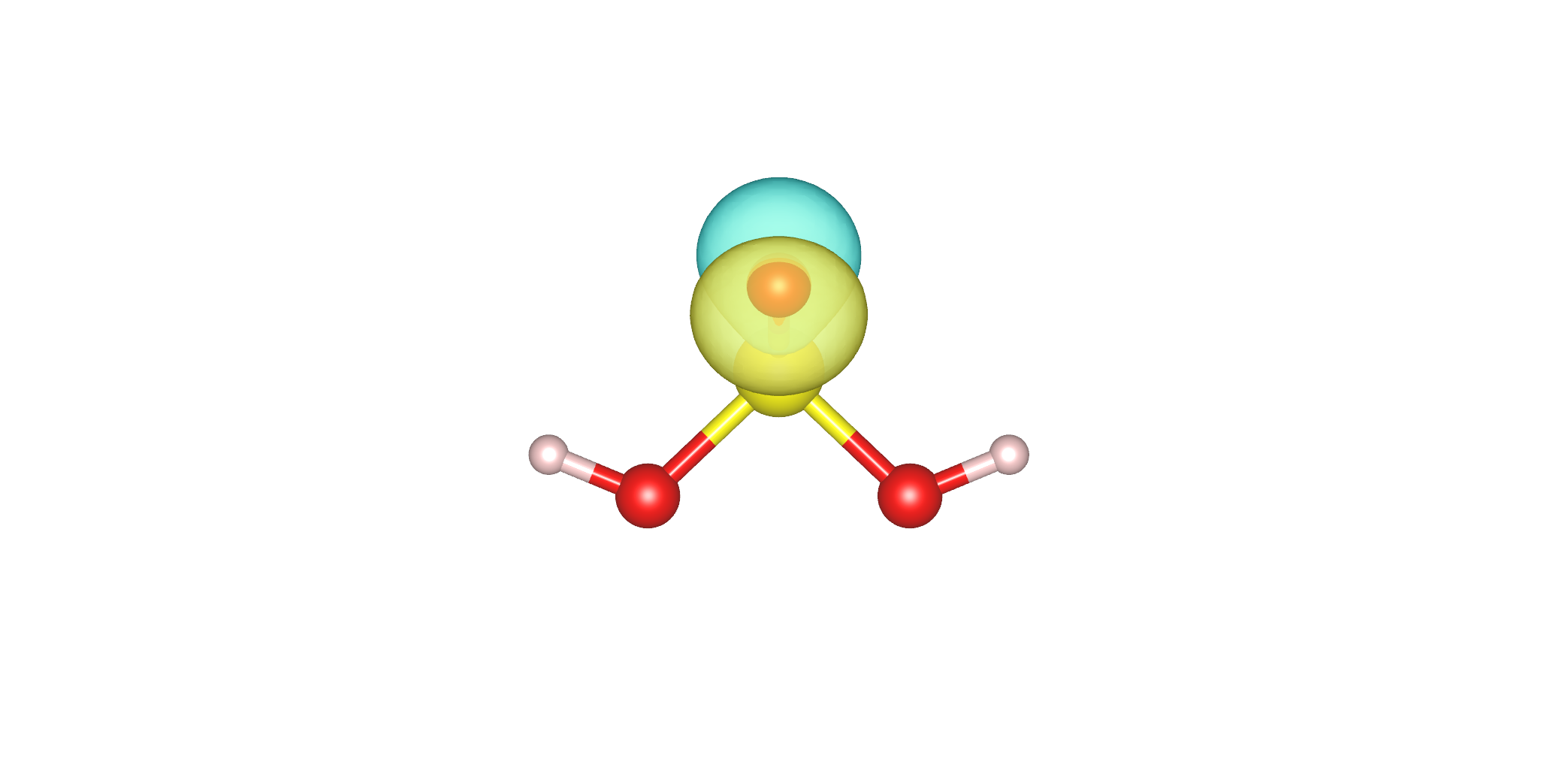}\\[-0.8em]
		\footnotesize (b2)
	\end{minipage}
	\begin{minipage}{0.15\linewidth}
		\centering
		\includegraphics[width=0.8\linewidth]{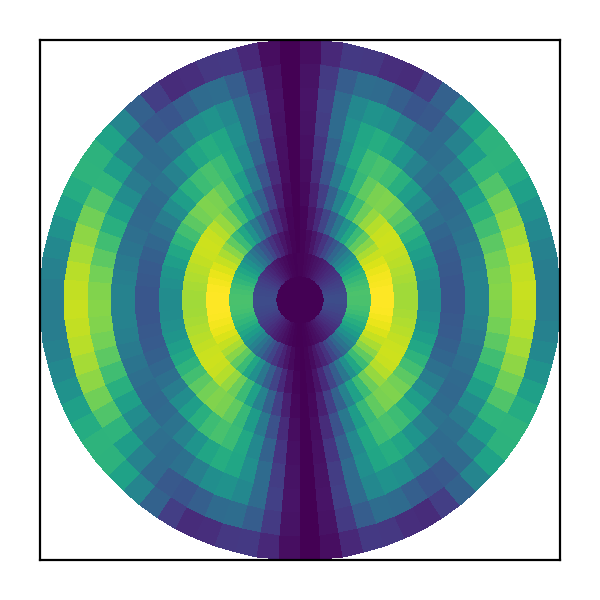}\\[-0.8em]
		\footnotesize (b3)
	\end{minipage}
	\begin{minipage}{0.15\linewidth}
		\centering
		\includegraphics[width=0.8\linewidth]{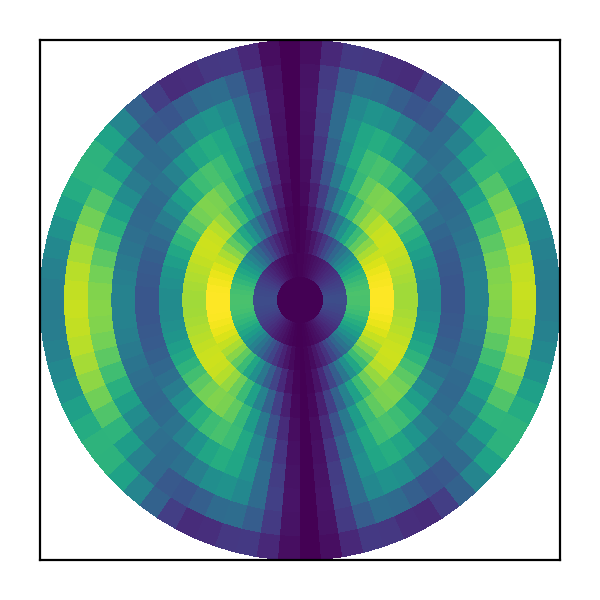}\\[-0.8em]
		\footnotesize (b4)
	\end{minipage}
	\begin{minipage}{0.15\linewidth}
		\centering
		\includegraphics[width=1\linewidth]{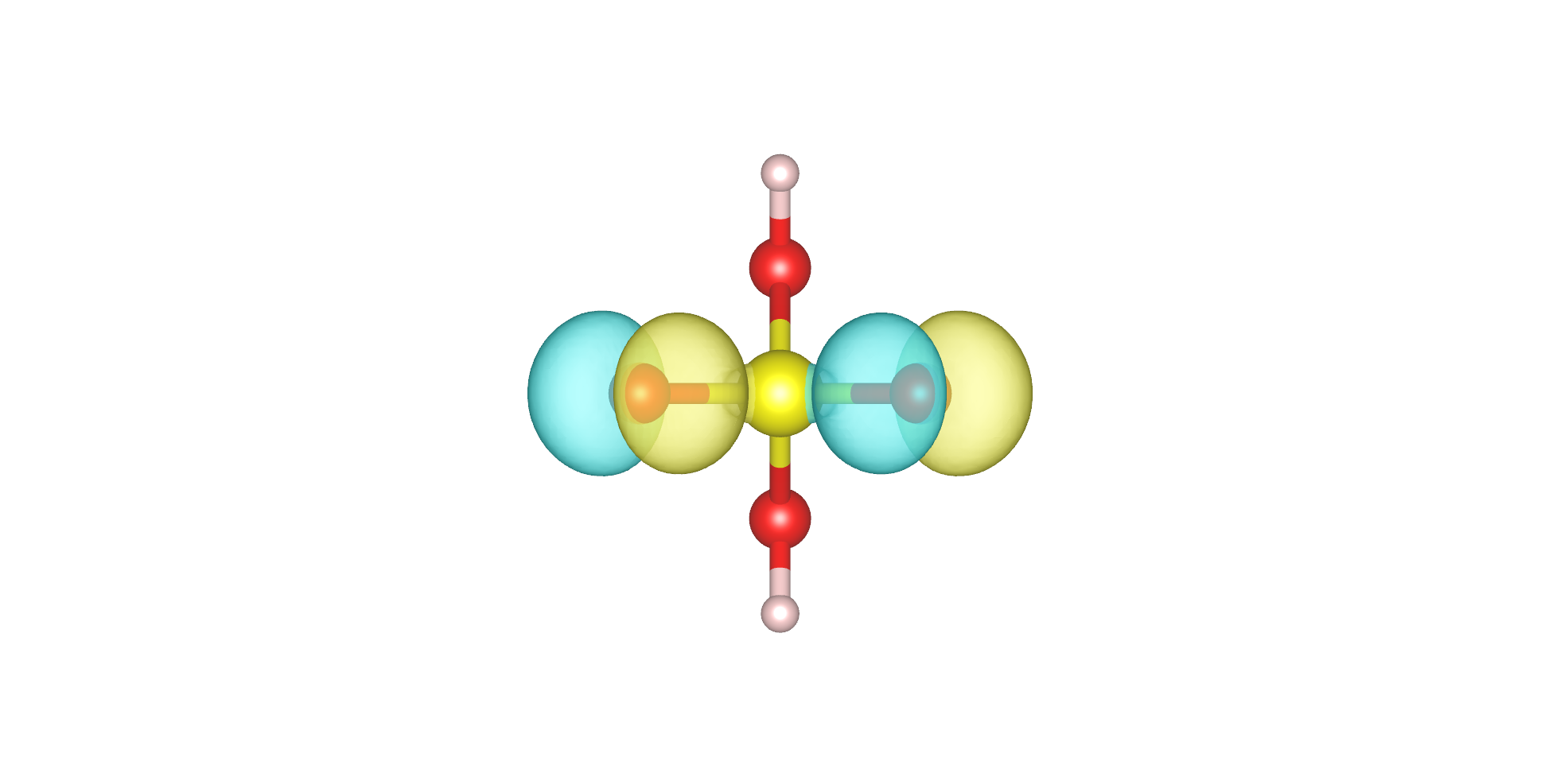}\\[-0.8em]
		\footnotesize (b5)
	\end{minipage}
    \begin{minipage}{0.15\linewidth}
		\centering
		\includegraphics[width=1\linewidth]{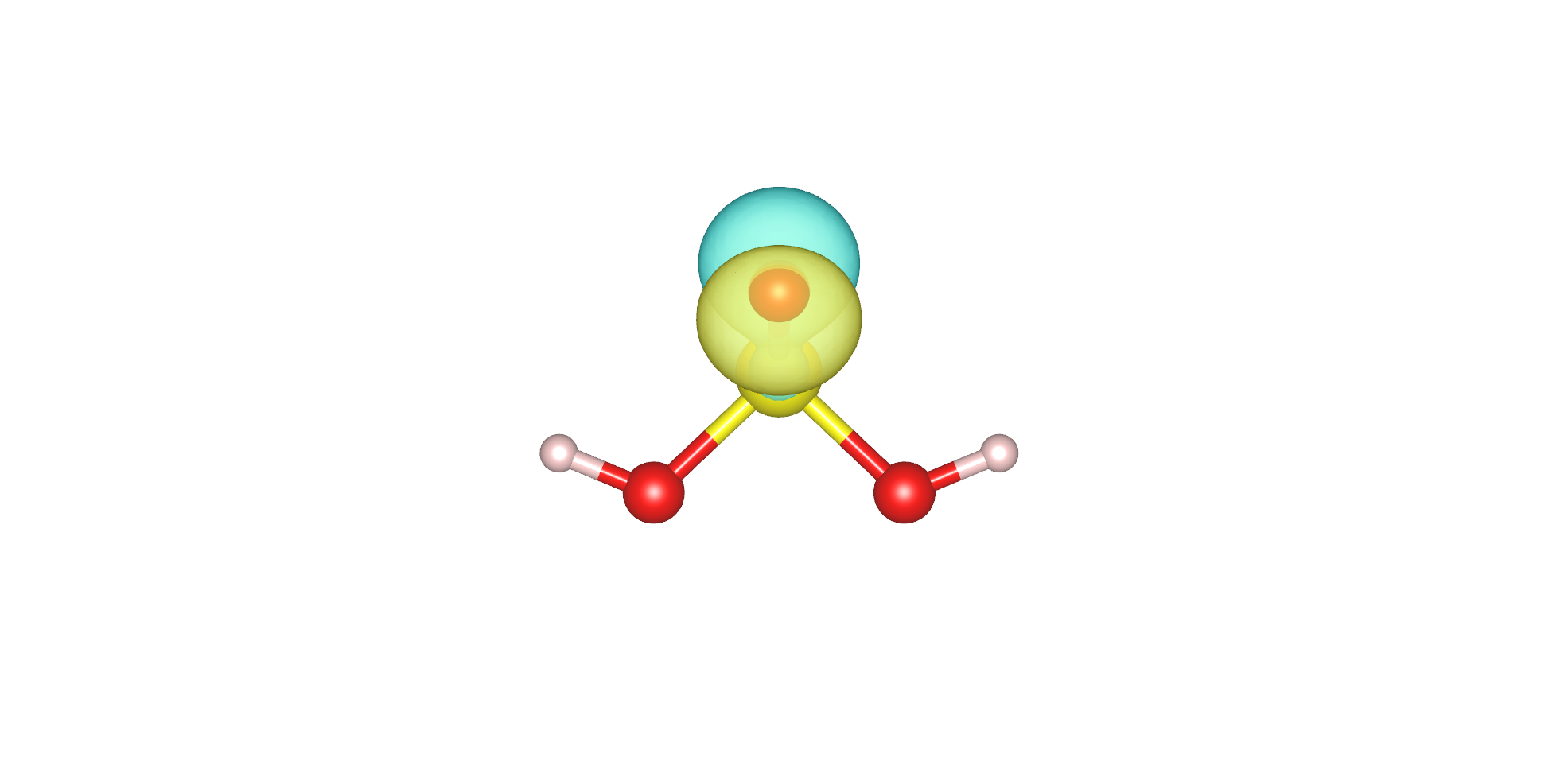}\\[-0.8em]
		\footnotesize (b6)
	\end{minipage}
    \\
	\begin{minipage}{0.15\linewidth}
		\centering
		\includegraphics[width=1\linewidth]{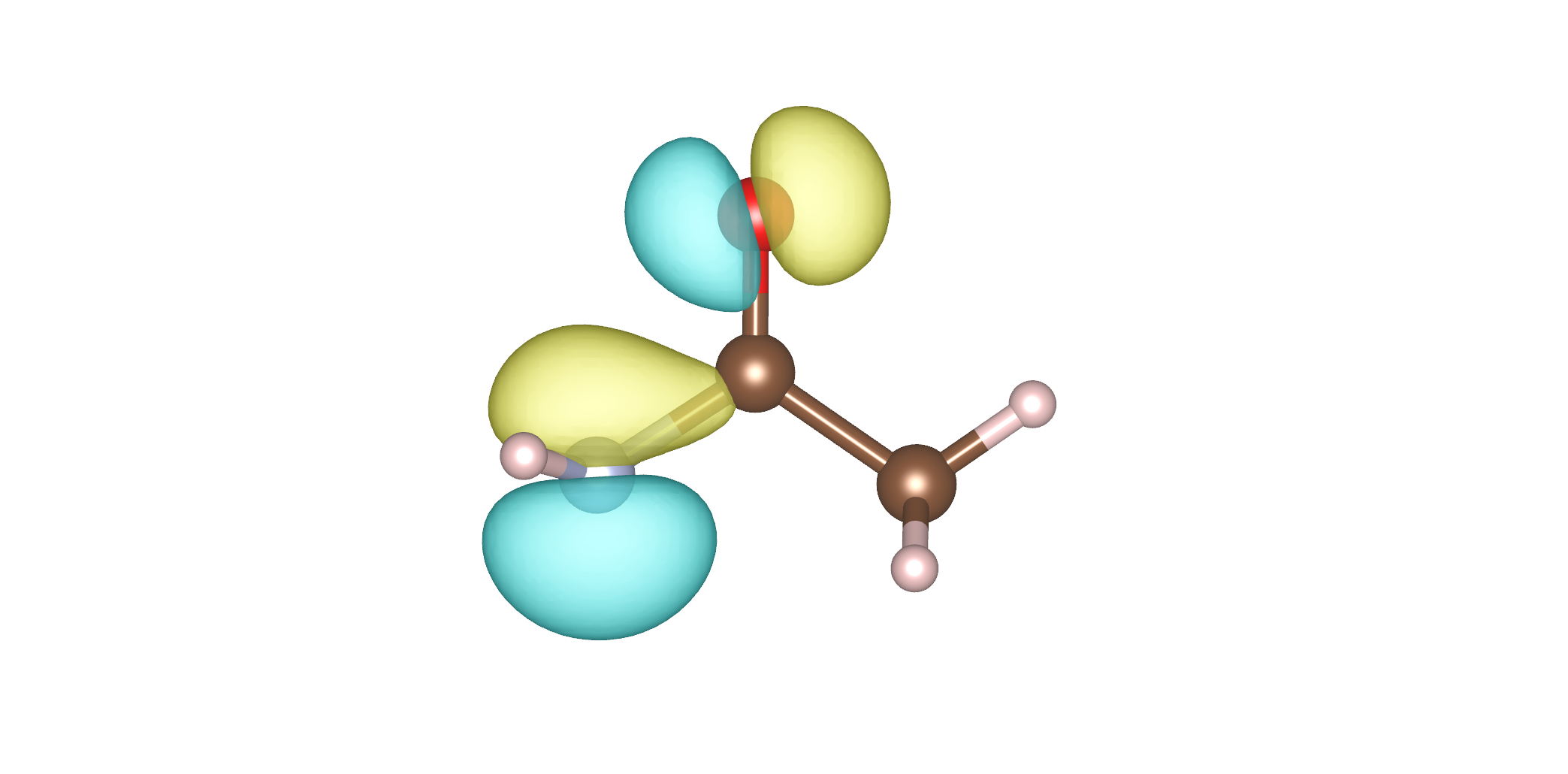}\\[-0.8em]
		\footnotesize (c1)
	\end{minipage}
	\begin{minipage}{0.15\linewidth}
		\centering
		\includegraphics[width=1\linewidth]{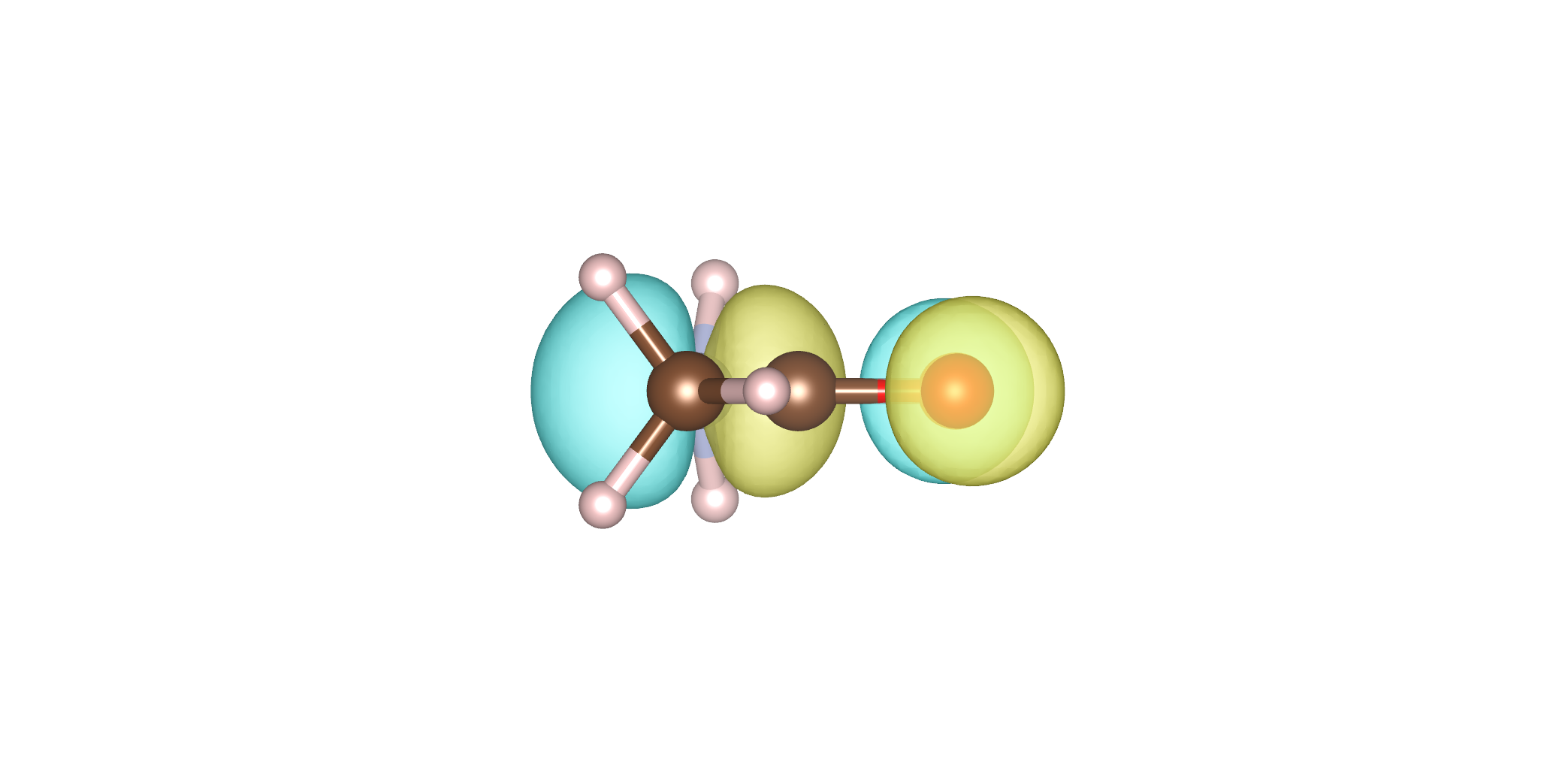}\\[-0.8em]
		\footnotesize (c2)
	\end{minipage}
	\begin{minipage}{0.15\linewidth}
		\centering
		\includegraphics[width=0.8\linewidth]{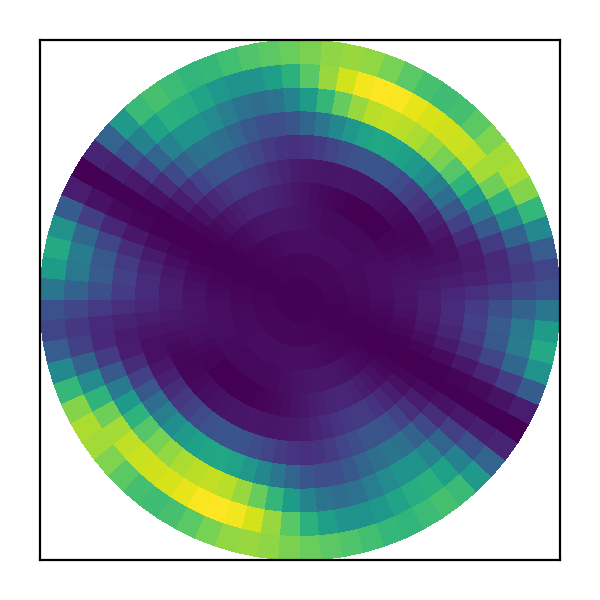}\\[-0.8em]
		\footnotesize (c3)
	\end{minipage}
	\begin{minipage}{0.15\linewidth}
		\centering
		\includegraphics[width=0.8\linewidth]{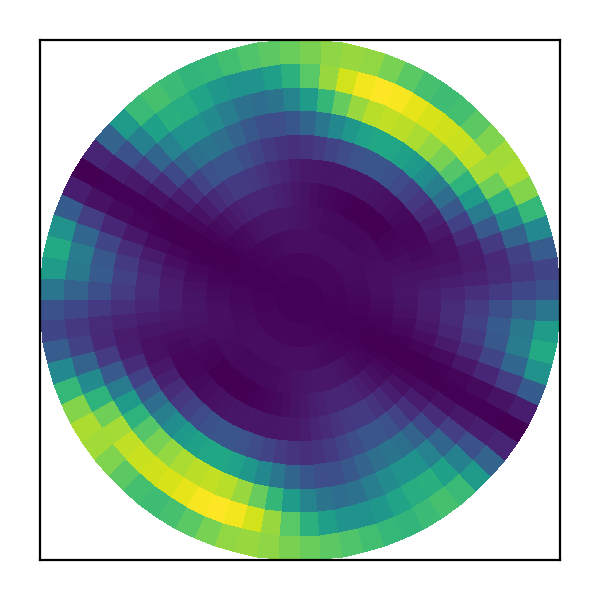}\\[-0.8em]
		\footnotesize (c4)
	\end{minipage}	
    \begin{minipage}{0.15\linewidth}
		\centering
		\includegraphics[width=1\linewidth]{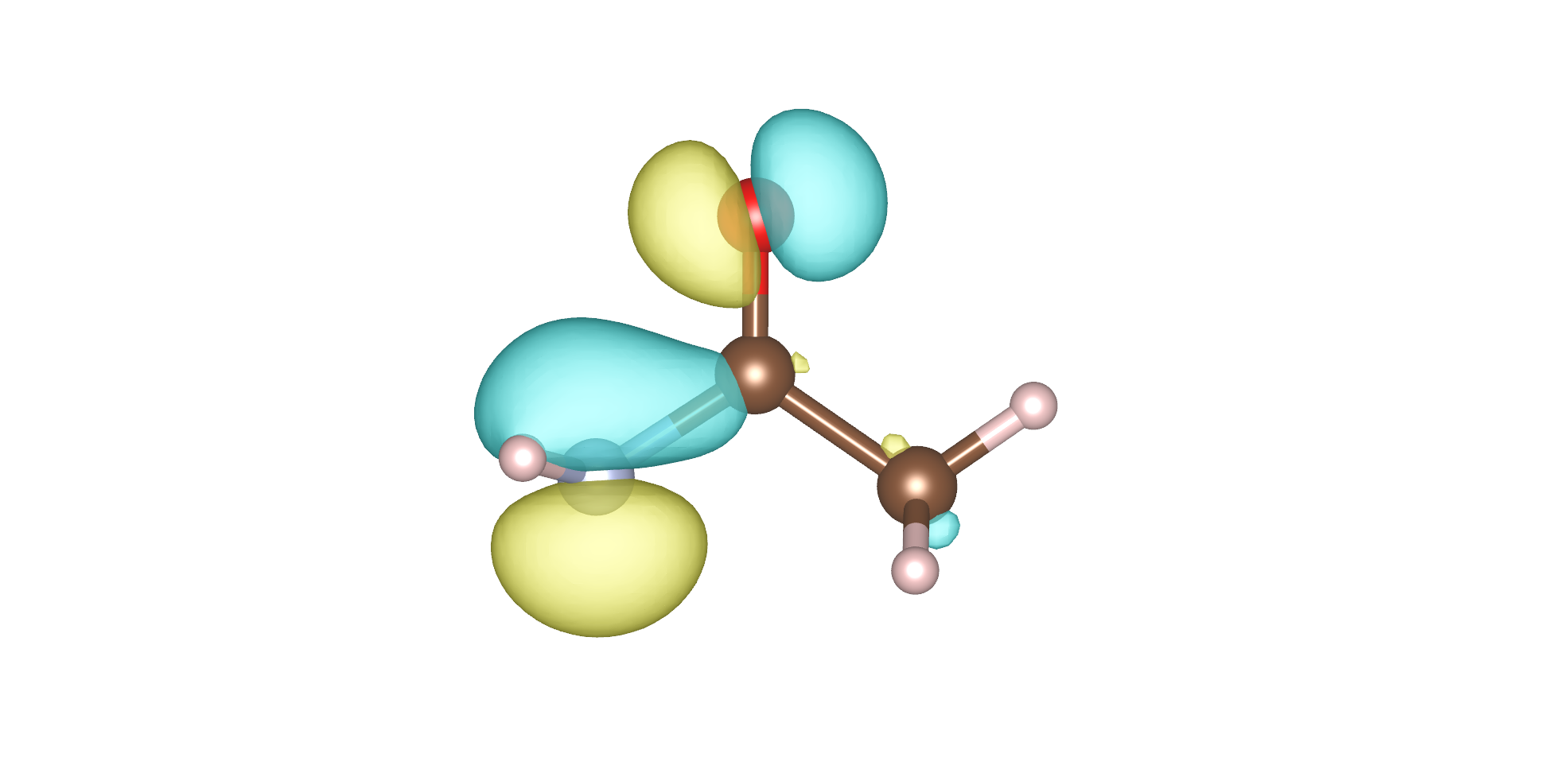}\\[-0.8em]
		\footnotesize (c5)
	\end{minipage}
    \begin{minipage}{0.15\linewidth}
		\centering
		\includegraphics[width=1\linewidth]{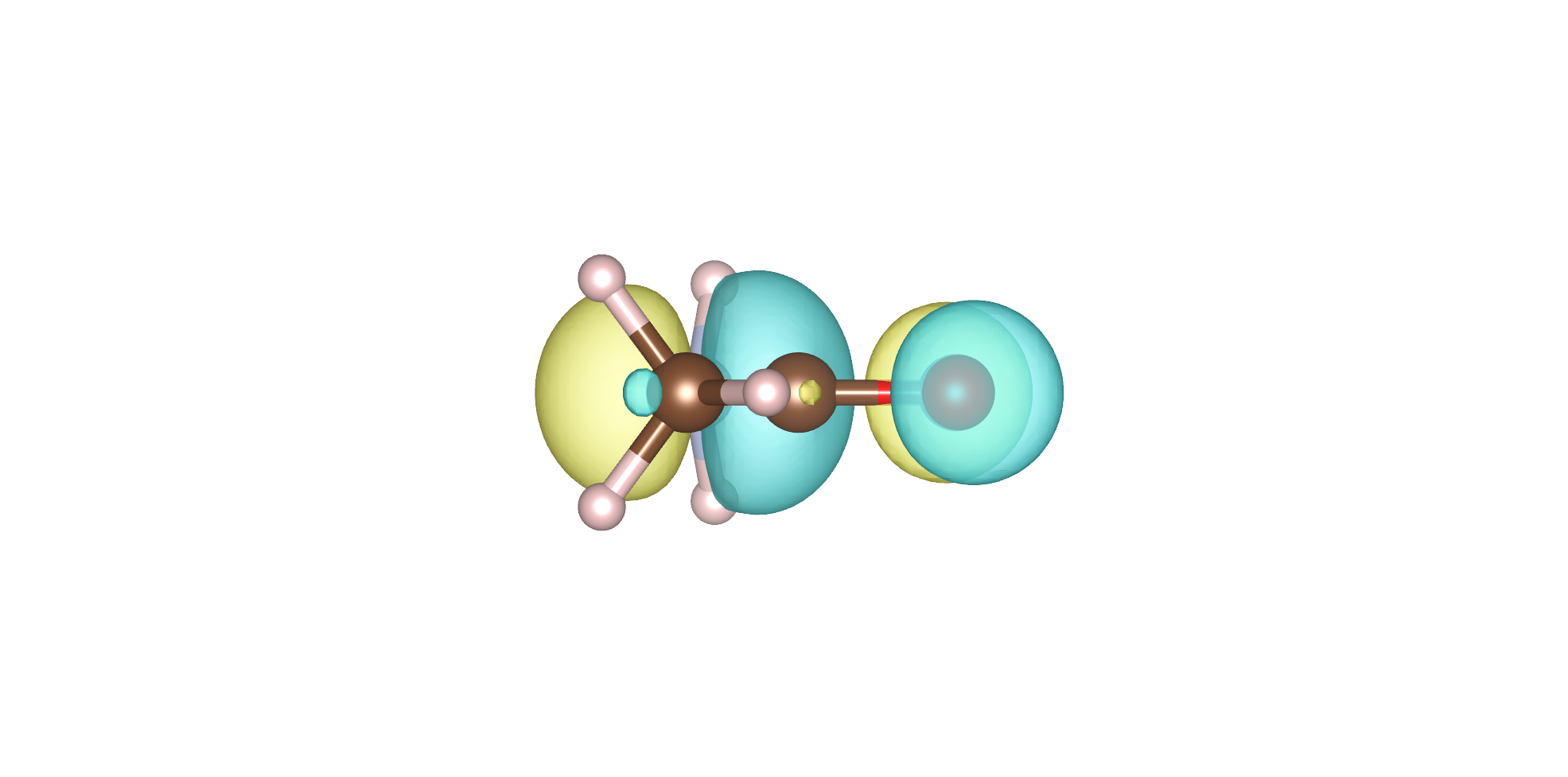}\\[-0.8em]
		\footnotesize (c6)
	\end{minipage}
	\\
	\begin{minipage}{0.15\linewidth}
		\centering
		\includegraphics[width=1\linewidth]{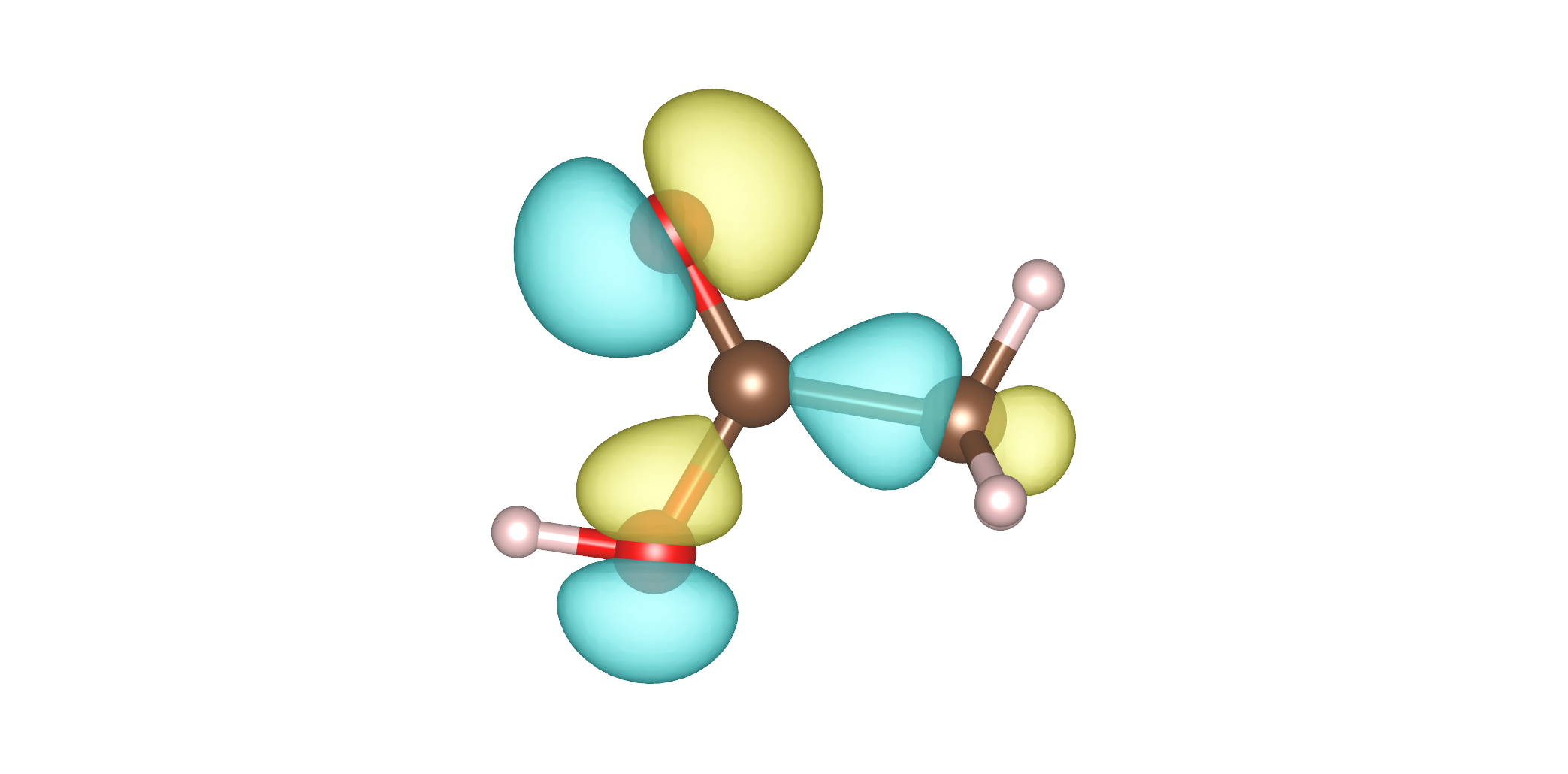}\\[-0.8em]
		\footnotesize (d1)
    \end{minipage}
    \begin{minipage}{0.15\linewidth}
		\centering
		\includegraphics[width=1\linewidth]{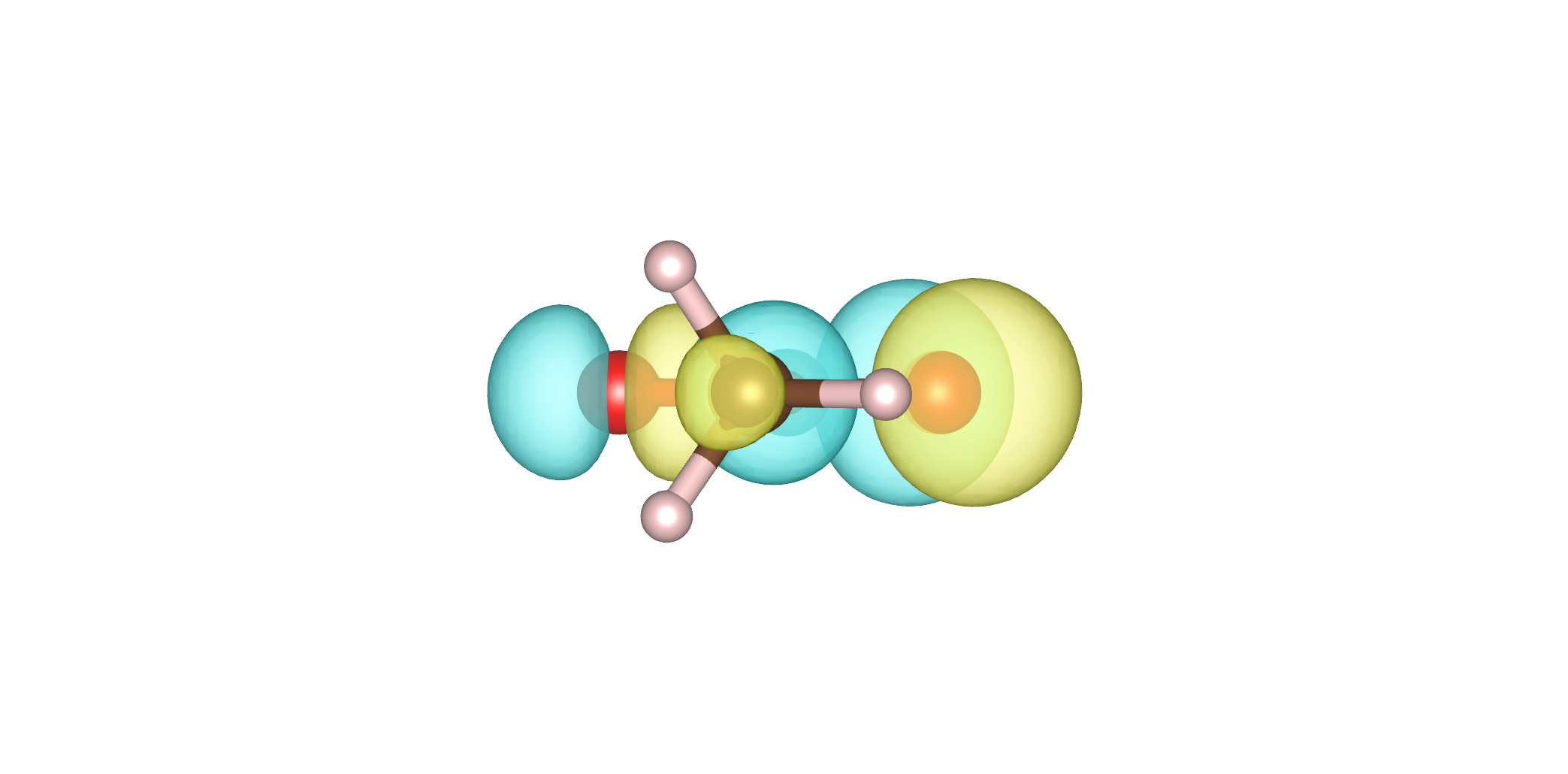}\\[-0.8em]
		\footnotesize (d2)
	\end{minipage}
	\begin{minipage}{0.15\linewidth}
		\centering
		\includegraphics[width=0.8\linewidth]{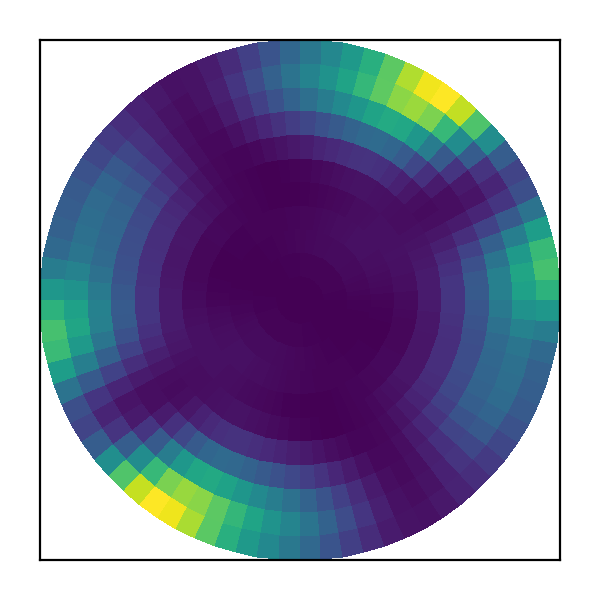}\\[-0.8em]
		\footnotesize (d3)
	\end{minipage}
	\begin{minipage}{0.15\linewidth}
		\centering
		\includegraphics[width=0.8\linewidth]{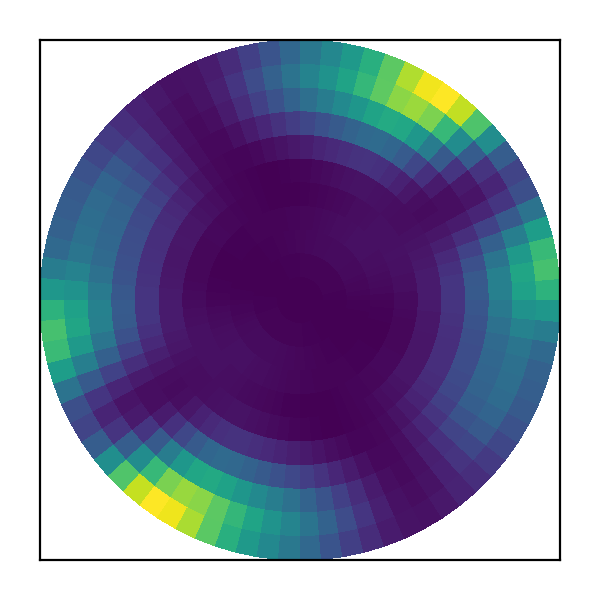}\\[-0.8em]
		\footnotesize (d4)
	\end{minipage}
	\begin{minipage}{0.15\linewidth}
		\centering
		\includegraphics[width=1\linewidth]{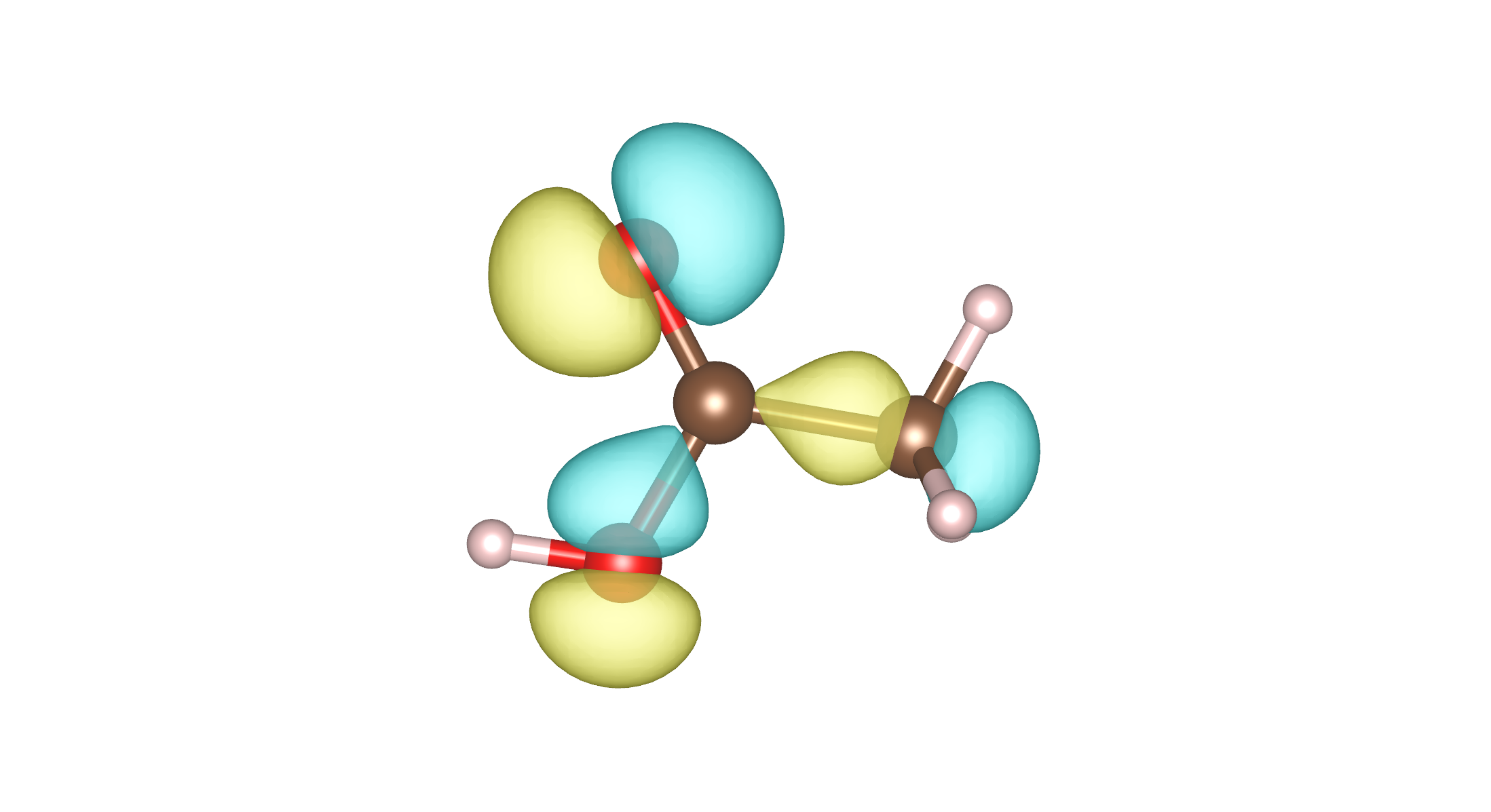}\\[-0.8em]
		\footnotesize (d5)
	\end{minipage}
    \begin{minipage}{0.15\linewidth}
		\centering
		\includegraphics[width=1\linewidth]{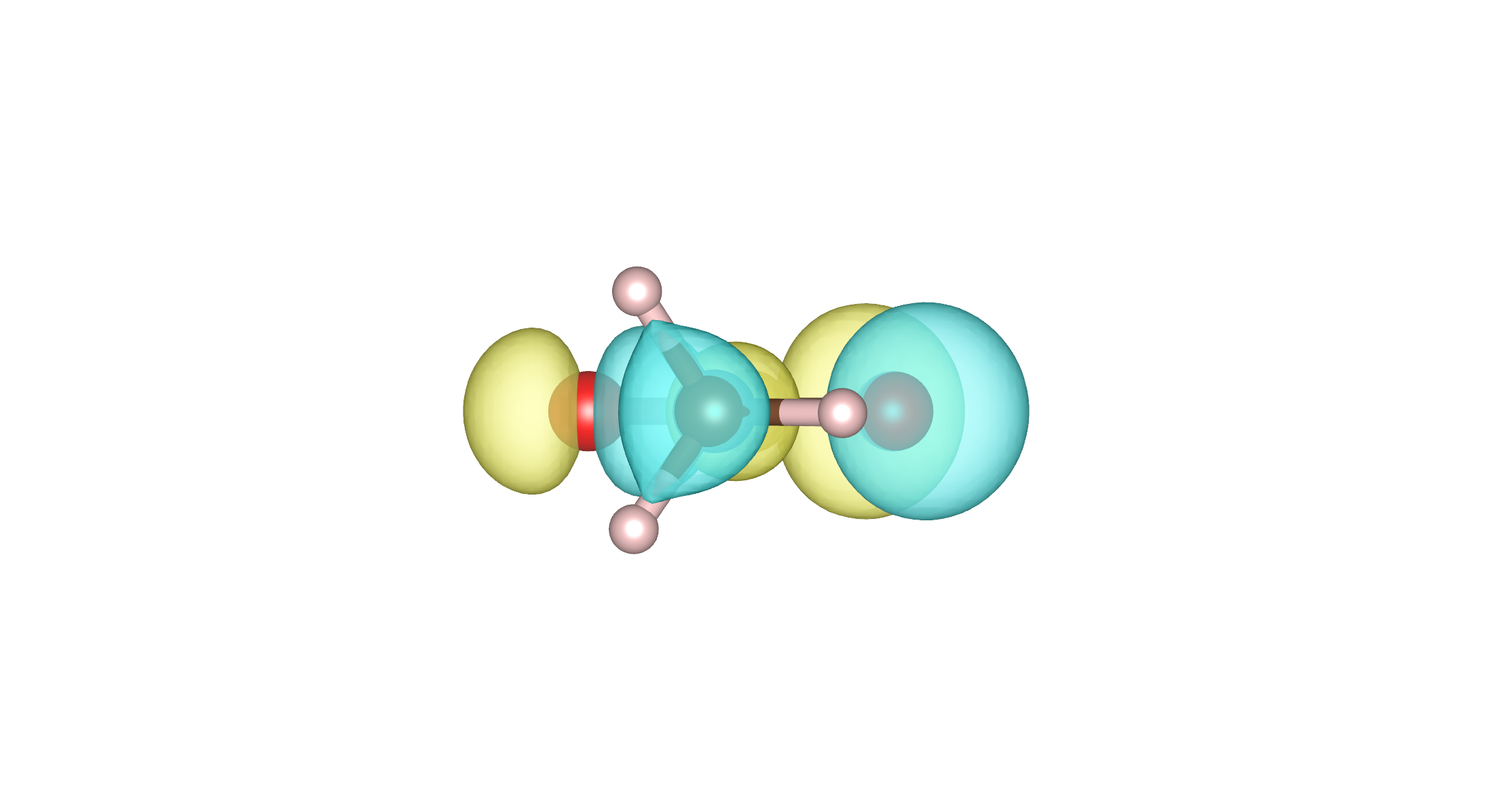}\\[-0.8em]
		\footnotesize (d6)
	\end{minipage}
    \\
    \begin{minipage}{0.15\linewidth}
		\centering
		\includegraphics[width=1\linewidth]{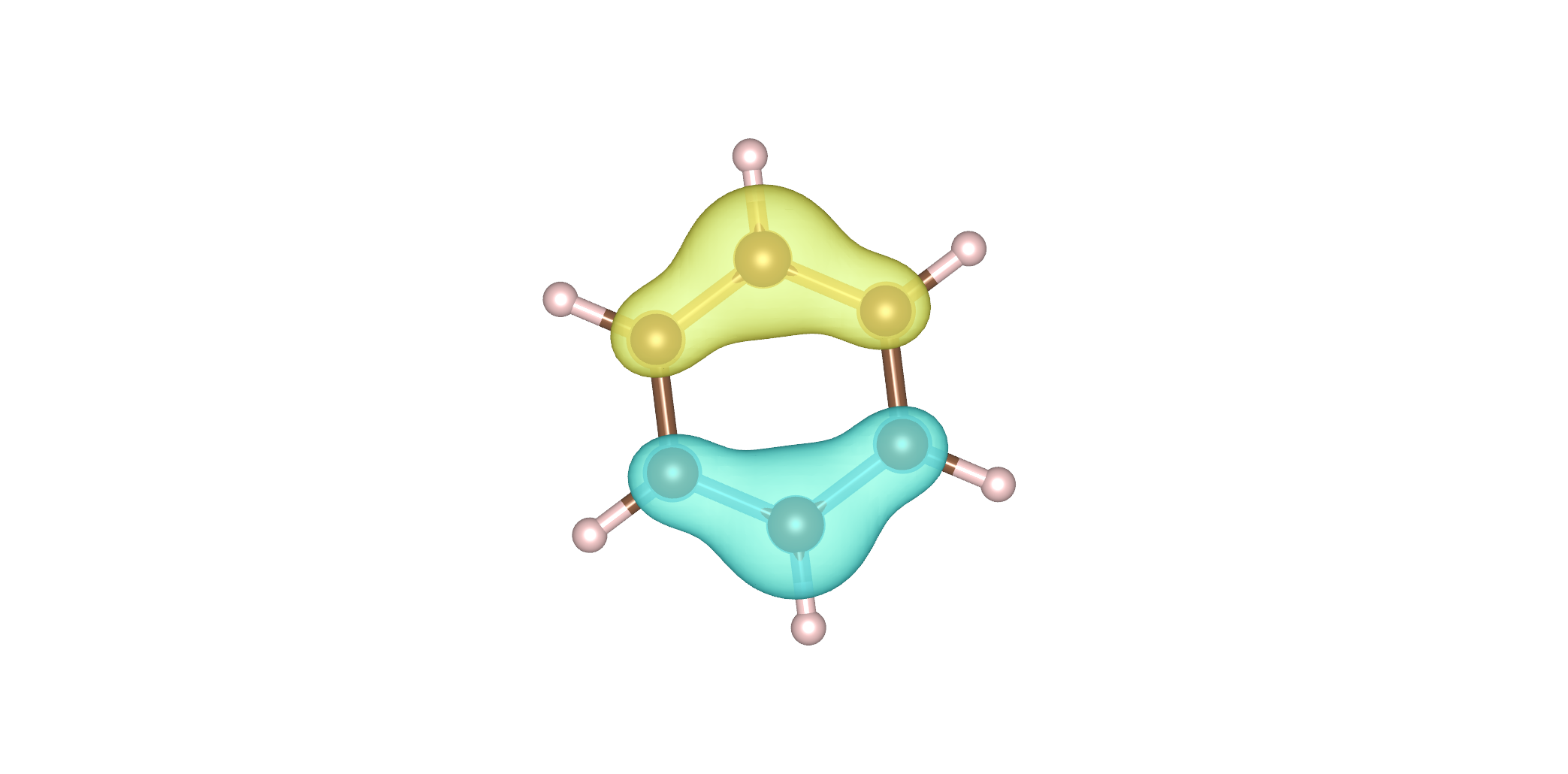}\\[-0.8em]
		\footnotesize (e1)
	\end{minipage}
    \begin{minipage}{0.15\linewidth}
	  \centering
		\includegraphics[width=1\linewidth]{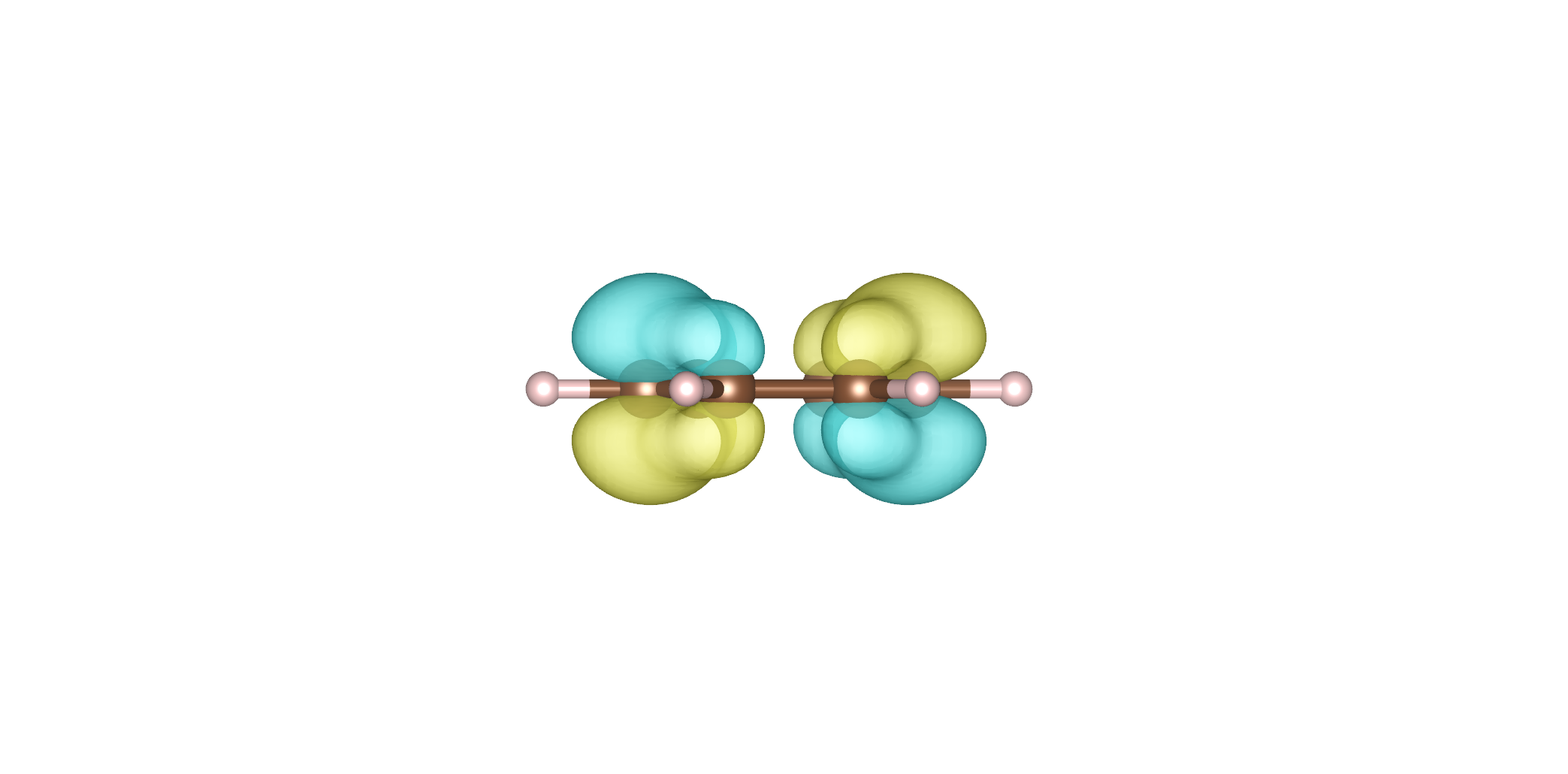}\\[-0.8em]
		\footnotesize (e2)
	\end{minipage}
	\begin{minipage}{0.15\linewidth}
		\centering
		\includegraphics[width=0.8\linewidth]{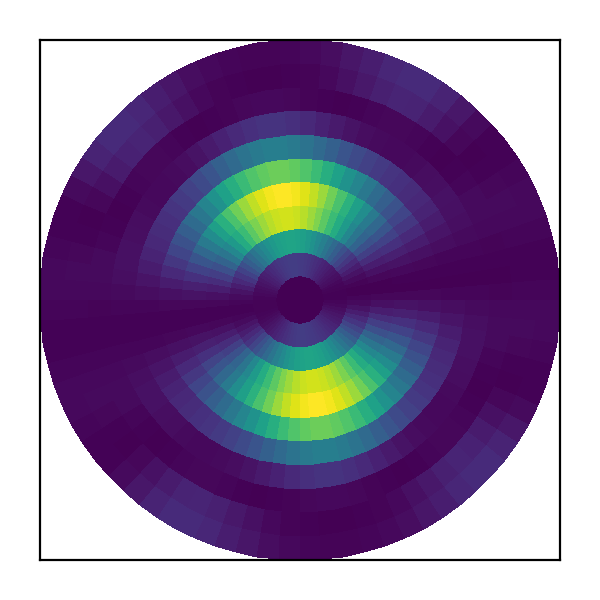}\\[-0.8em]
		\footnotesize (e3)
	\end{minipage}
	\begin{minipage}{0.15\linewidth}
		\centering
		\includegraphics[width=0.8\linewidth]{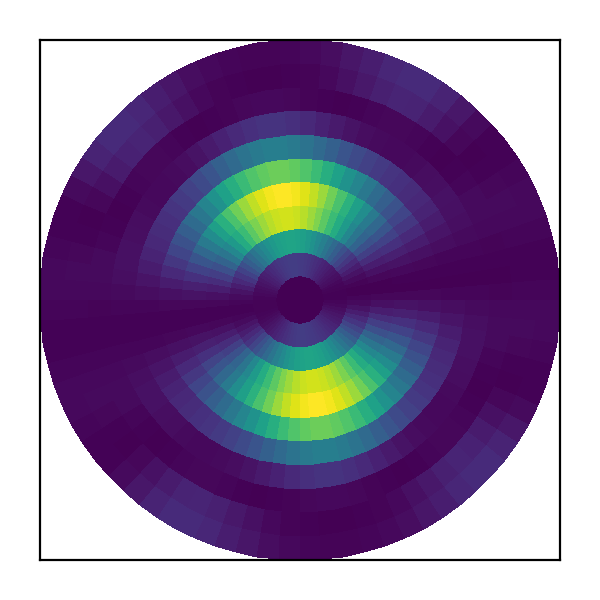}\\[-0.8em]
		\footnotesize (e4)
	\end{minipage}
	\begin{minipage}{0.15\linewidth}
		\centering
		\includegraphics[width=1\linewidth]{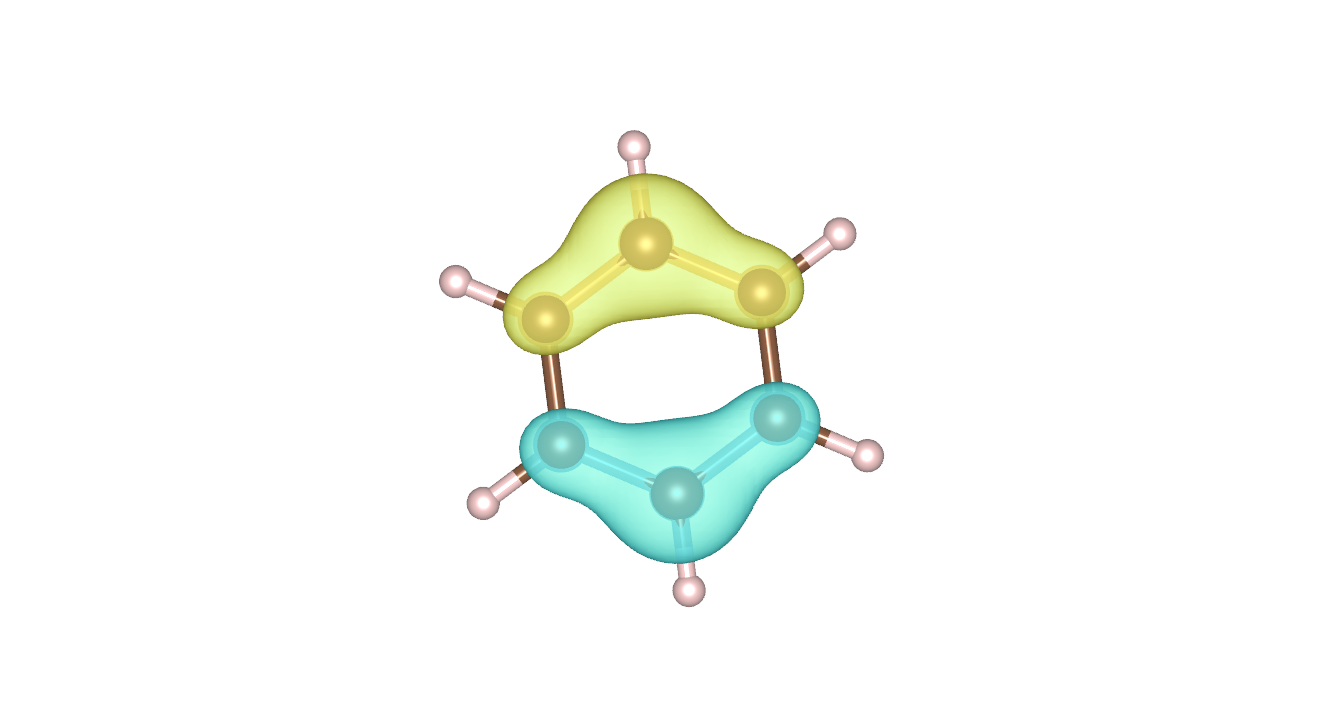}\\[-0.8em]
		\footnotesize (e5)
    \end{minipage}
    \begin{minipage}{0.15\linewidth}
		\centering
		\includegraphics[width=1\linewidth]{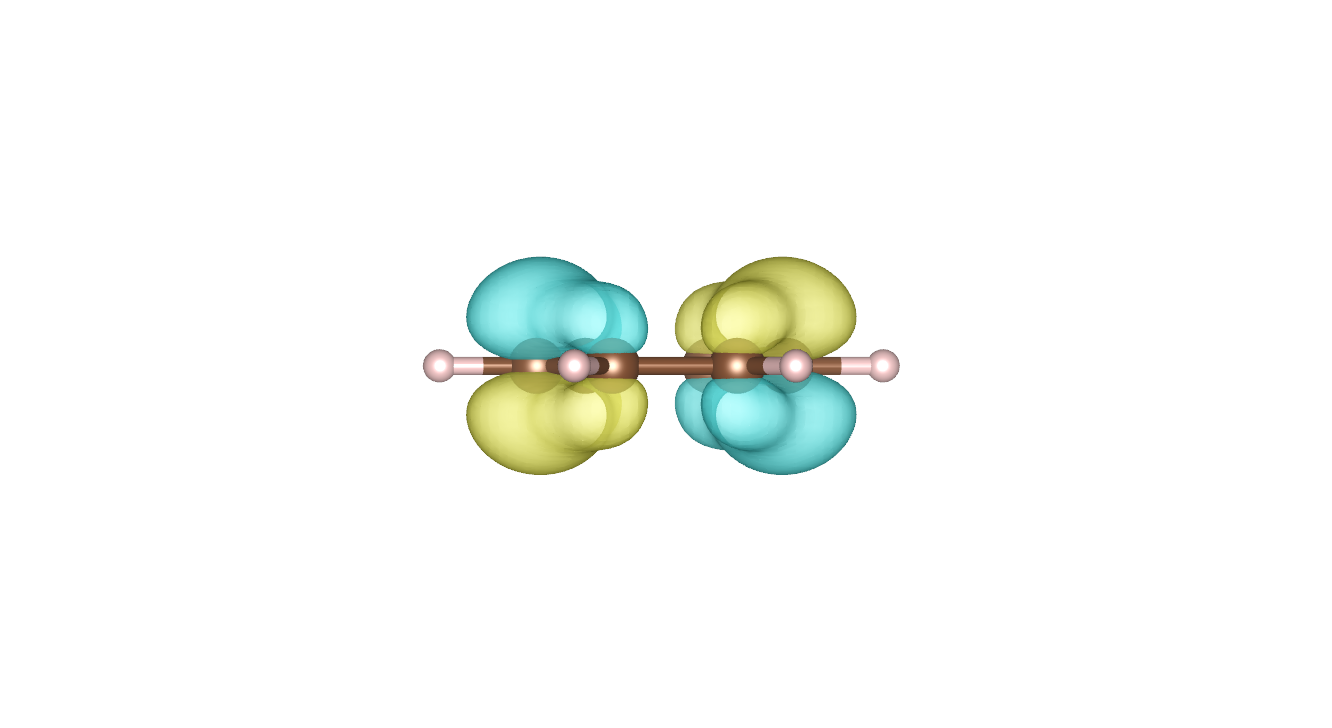}\\[-0.8em]
		\footnotesize (e6)
	\end{minipage}
    \\
	\begin{minipage}{0.15\linewidth}
		\centering
		\includegraphics[width=1\linewidth]{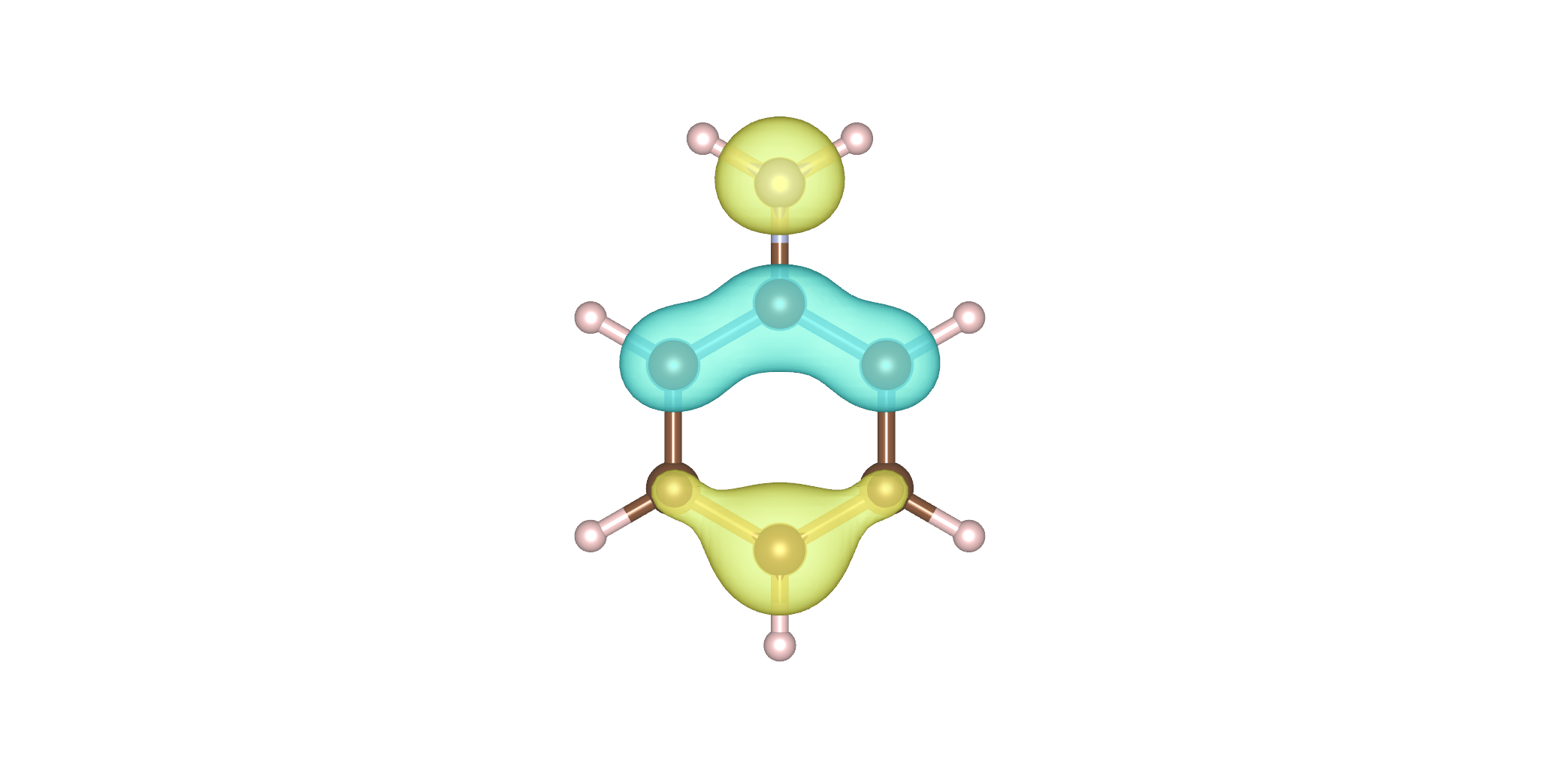}\\[-0.8em]
		\footnotesize (f1)
	\end{minipage}
    \begin{minipage}{0.15\linewidth}
	  \centering
		\includegraphics[width=1\linewidth]{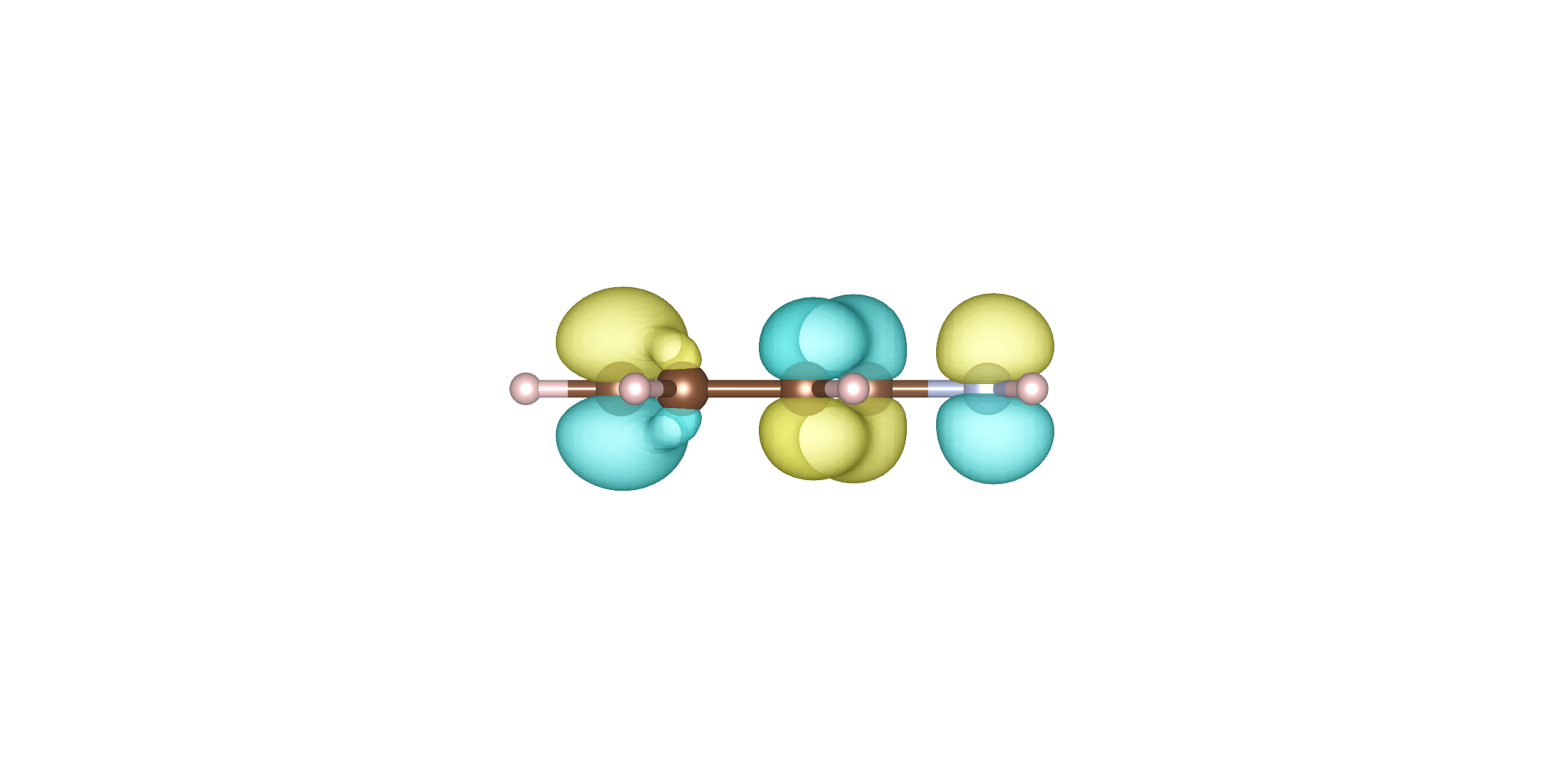}\\[-0.8em]
		\footnotesize (f2)
	\end{minipage}
	\begin{minipage}{0.15\linewidth}
		\centering
		\includegraphics[width=0.8\linewidth]{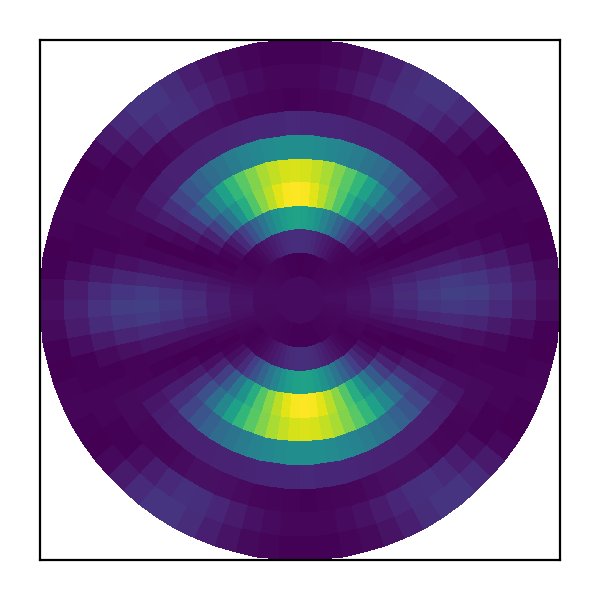}\\[-0.8em]
		\footnotesize (f3)
	\end{minipage}
	\begin{minipage}{0.15\linewidth}
		\centering
		\includegraphics[width=0.8\linewidth]{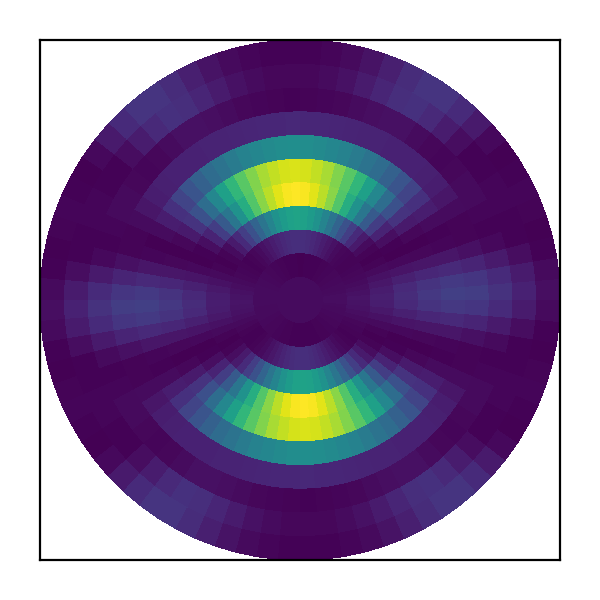}\\[-0.8em]
		\footnotesize (f4)
	\end{minipage}
	\begin{minipage}{0.15\linewidth}
		\centering
		\includegraphics[width=1\linewidth]{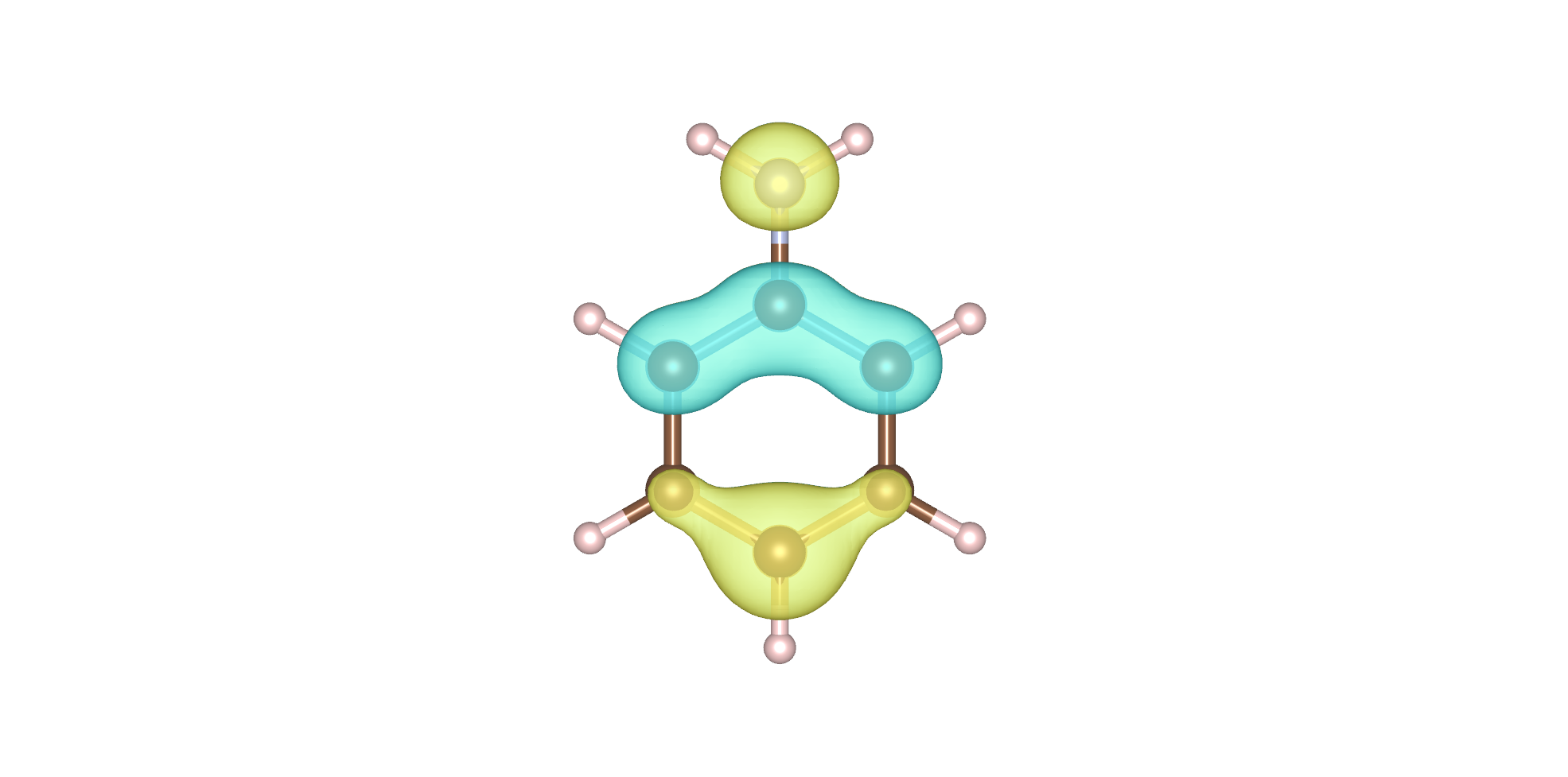}\\[-0.8em]
		\footnotesize (f5)
	\end{minipage}
    \begin{minipage}{0.15\linewidth}
	\centering
	\includegraphics[width=1\linewidth]{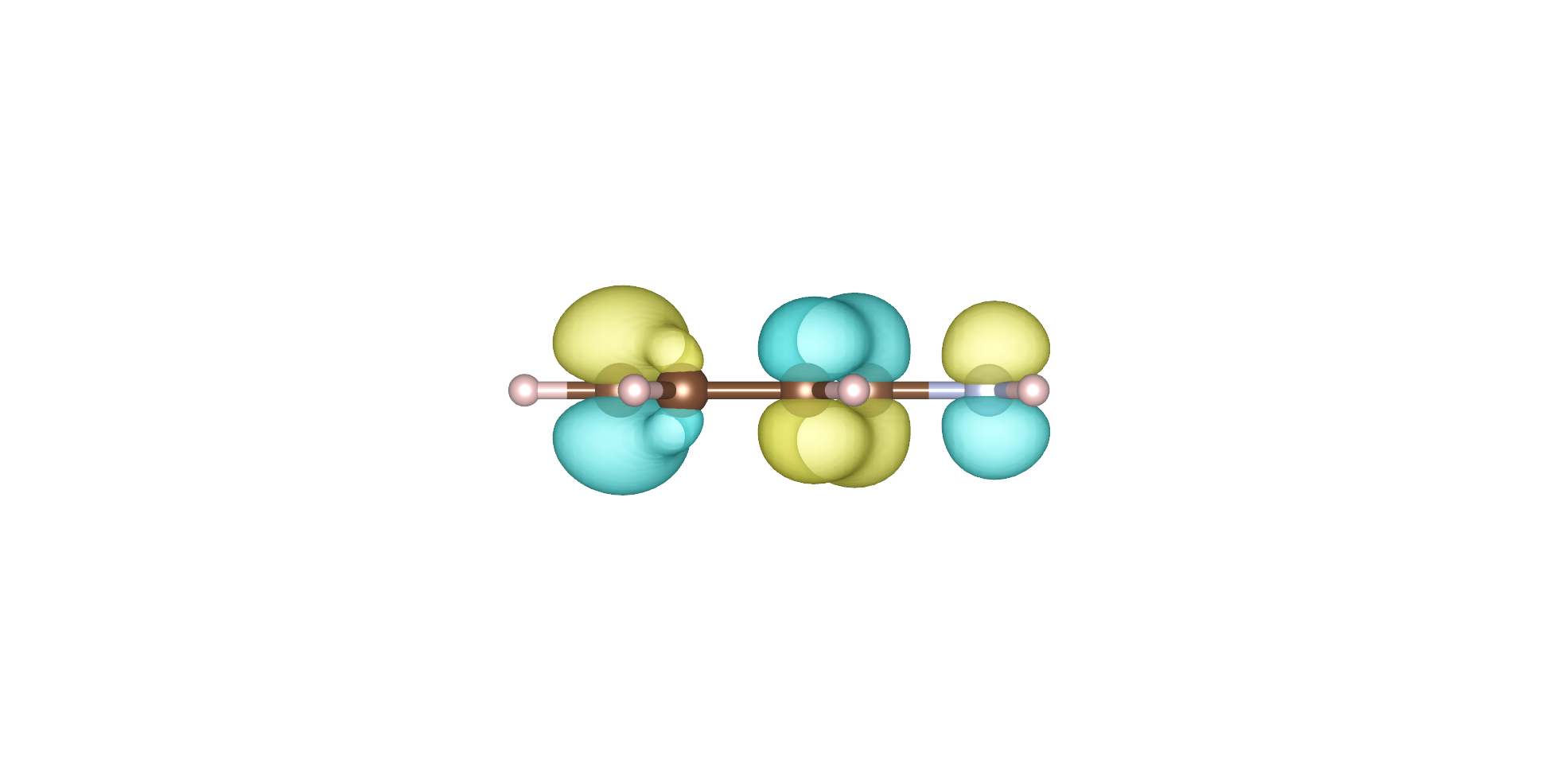}\\[-0.8em]
	\footnotesize (f6)
	\end{minipage}
	\\
	\caption{PMMs of some simple molecules and their molecule orbitals.		
	}
	\label{fig:simple-molecules}
\end{figure*}

\begin{figure*}[t]
	\centering
	\begin{minipage}{0.15\linewidth}
		\centering
		\includegraphics[width=1\linewidth]{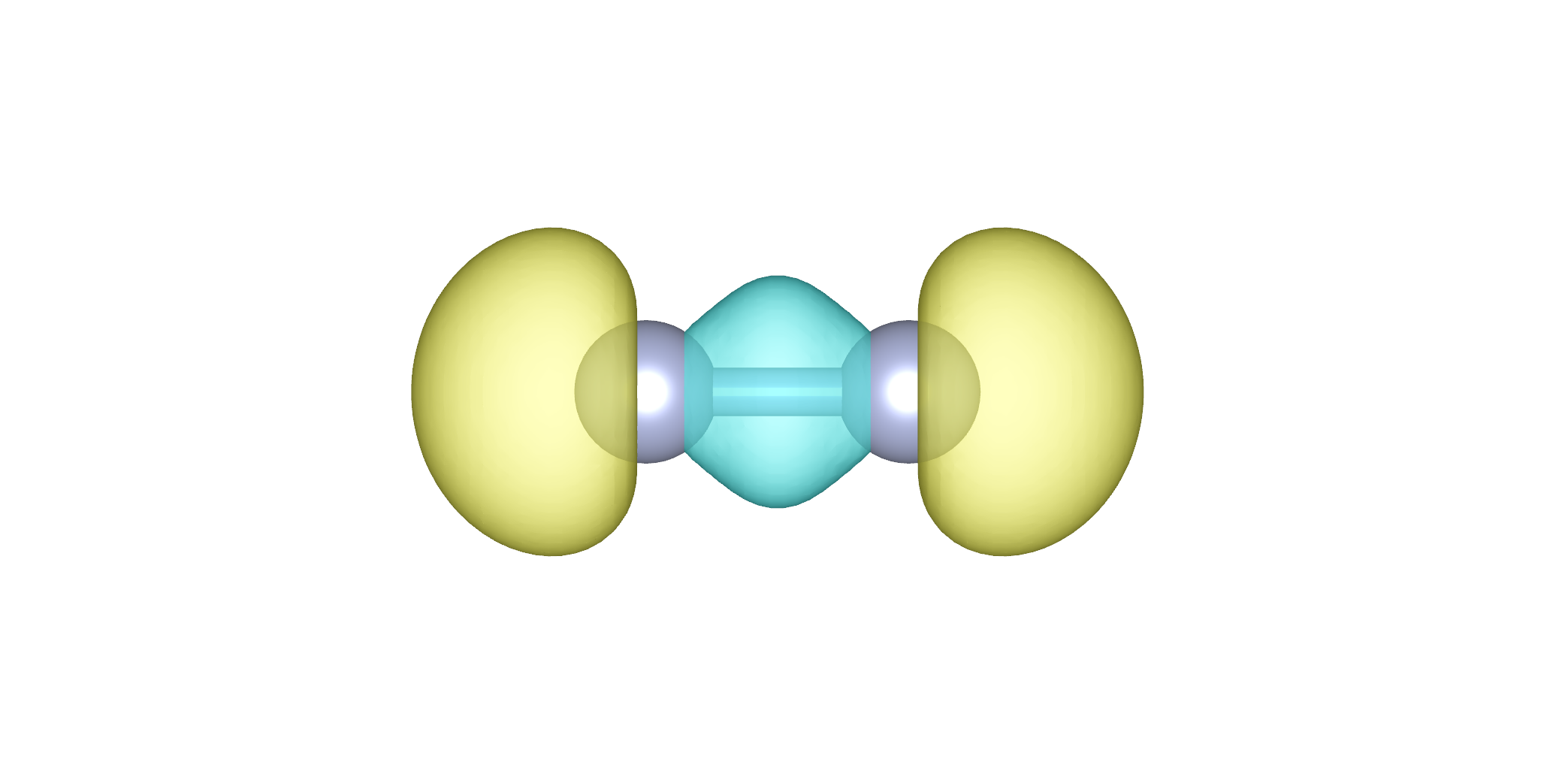}\\[-0.8em]
		\footnotesize (a1)
	\end{minipage}
    \begin{minipage}{0.15\linewidth}
		\centering
		\includegraphics[width=1\linewidth]{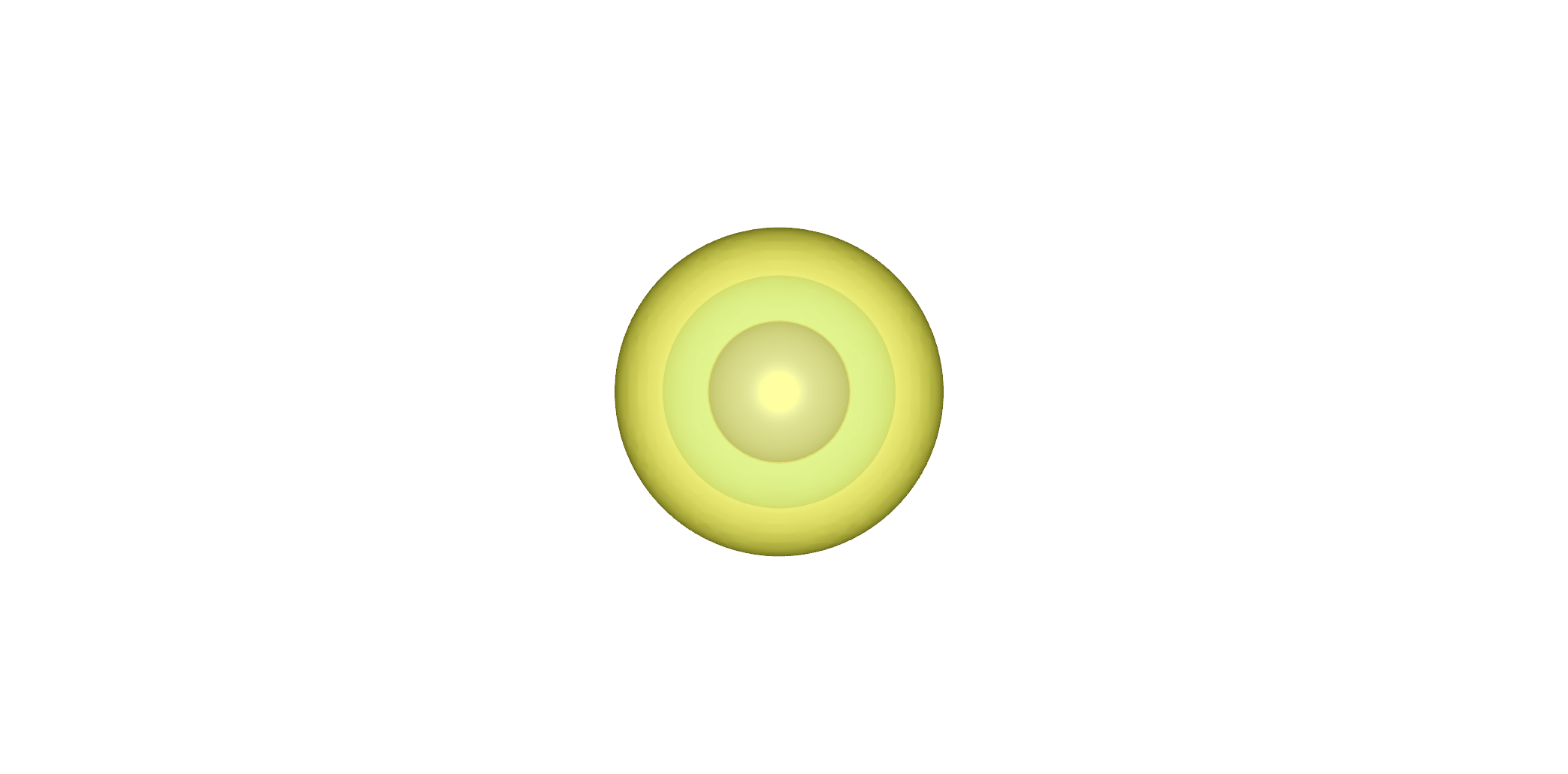}\\[-0.8em]
		\footnotesize (a2)
	\end{minipage}
	\begin{minipage}{0.15\linewidth}
		\centering
		\includegraphics[width=0.8\linewidth]{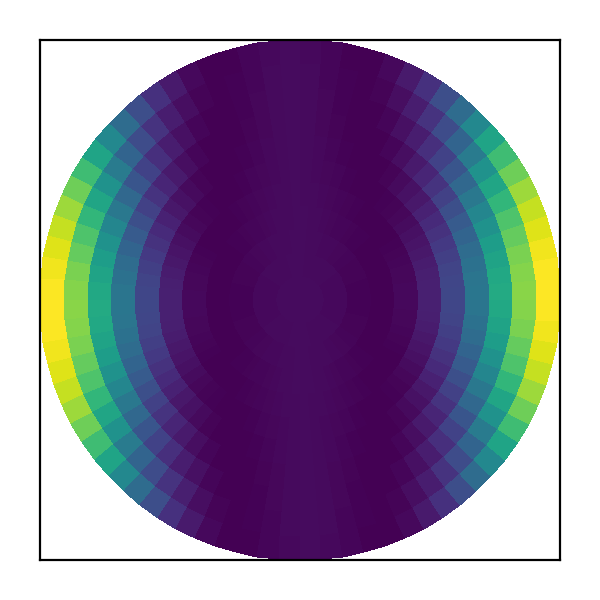}\\[-0.8em]
		\footnotesize (a3)
	\end{minipage}
	\begin{minipage}{0.15\linewidth}
		\centering
		\includegraphics[width=0.8\linewidth]{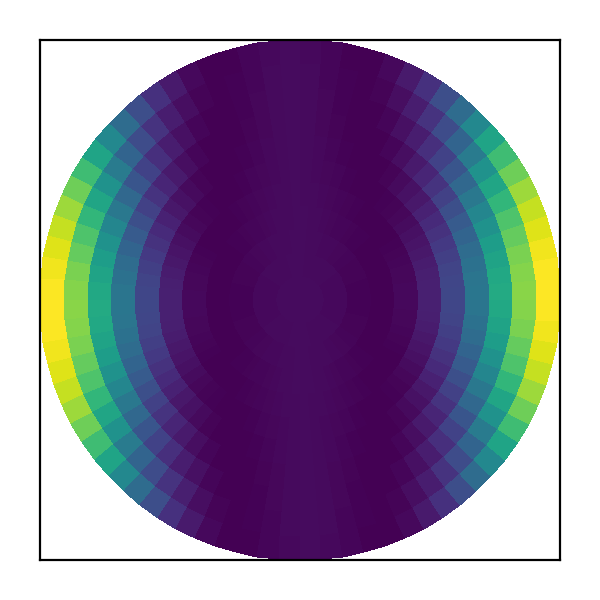}\\[-0.8em]
		\footnotesize (a4)
	\end{minipage}
	\begin{minipage}{0.15\linewidth}
		\centering
		\includegraphics[width=1\linewidth]{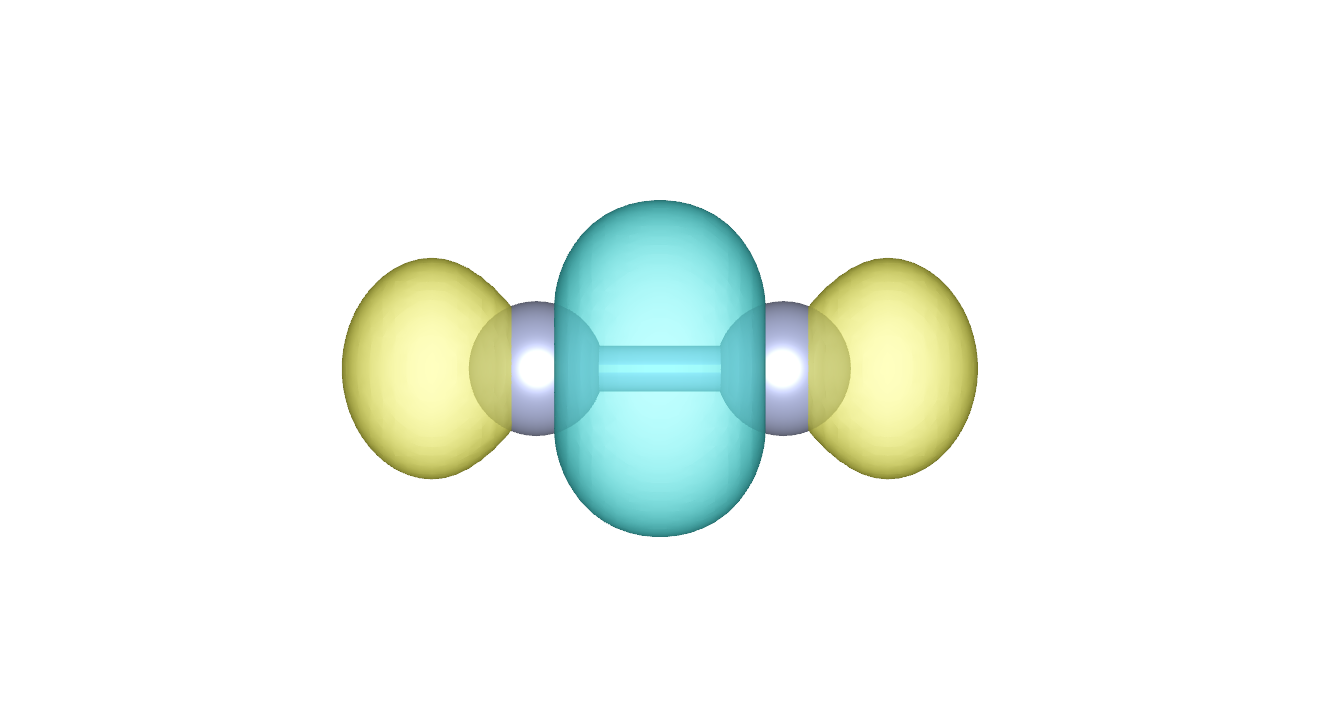}\\[-0.8em]
		\footnotesize (a5)
    \end{minipage}
    \begin{minipage}{0.15\linewidth}
		\centering
		\includegraphics[width=1\linewidth]{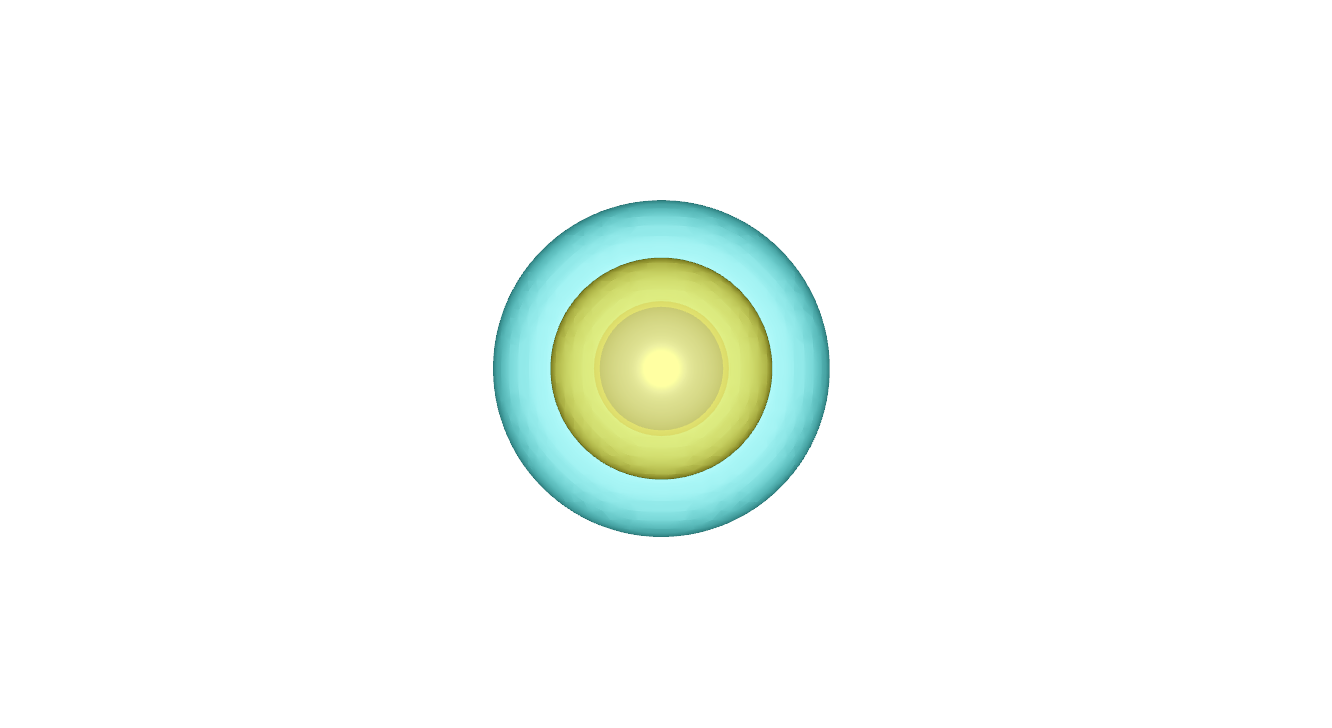}\\[-0.8em]
		\footnotesize (a6)
	\end{minipage}
	\\
	\caption{
		PMMs of some simple molecules and their molecule orbitals.	}
	\label{fig:simple-molecules}
\end{figure*}

\subsection{Phase retrieval from noisy PMMs using sparse PhaseLift}
\par
We next show that our method can estimate the molecular orbital even from ideal experimeltal PMMs. Firstly, we explain the method to generate experimental PMM on a computer. 
\par
All elements in a PMM $\vec{z}=(z_1, \cdots, z_M)^\top$ are non-negative and $\vec{z}$ is normalized as $\|\vec{z}\|_1=1$, where $\|\vec{z}\|_1 =\sum_{m=1}^M |z_m|$ denotes the $\ell_1$ norm of a vector $\vec{z}$.  Therefore, a PMM $\vec{z}$ can be regarded as a probability vector. We define a PMM with error as a histogram of samples drawn from the probability vector $\vec{z}$. The detail is as follows. Let $s_1, \cdots, s_T$ be samples those are independently drawn from a uniform distribution on the real interval $[0,1)$, where $T$ is the number of samples, which is called the \textit{total photoelectron number}. Let $T_1 =|\{t : s_t < z_1 \}|$ be the number of samples those are less than $z_1$. For $m \in \{2, \cdots, M\}$, let $T_m = |\{t : z_1 + \cdots + z_{m-1} \le s_t < z_1 + \cdots + z_m \}|$  be the number of samples those are greater than or equal to $z_1 + \cdots + z_{m-1}$ and less than $z_1 + \cdots + z_m$. Note that $T_1 + \cdots + T_m = T$. A PMM with error $\vec{z}$ is defined by 
\begin{linenomath}
\begin{align}
	\vec{z}\,'= \biggl( \frac{T_1}T, \cdots, \frac{T_m}T \biggr)^\top, 
\end{align}
\end{linenomath}
which is referred to as a \textit{quasi PMM}. This process can be regarded as a simple model for the observation process of PMMs. 
 \par
Using our method, we performed the molecular orbital estimation from the noisy PMM (including noise). The theoretical PMM of F4-TCNQ at $T=10^3$ (Fig. 4(b)) appears rough PMM. On the other hand, at $T=10^4$ and $10^5$ (Figs. 4(c) and 4(d)) seem to be smooth PMMs similar to the PMM without noise (Fig. 4(a)). They were increasing the value $T$ yields decreasing TVD.
\par
MO-error is plotted against the TVD for each $T$ as shown in figure 4(e). The red ($T=10^3$), green ($T=10^4$), and blue ($T=10^5$) dots clearly show the relationship between the TVD and the MO-error were increasing the value T yields decreasing MO-error. Even if TVD is $T=10^4$, the MO-error is as small as 0.02, indicating that our method can estimate under noise. This value is larger than the MO-error based on the noiseless PMM (Fig. 4(a)). However, it is comparable to the MO-error of acetamide and aceticacid based on the noiseless PMMs.

\begin{figure}[t]
	\centering
	\begin{minipage}{0.3\linewidth}
		\centering \includegraphics[width=1\linewidth]{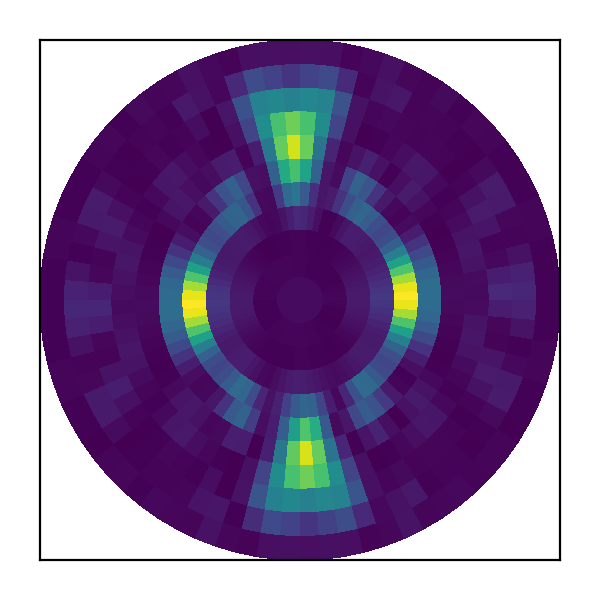}\\[-0.0em]
		\footnotesize (a)
	\end{minipage}
	\begin{minipage}{0.3\linewidth}
		\centering \includegraphics[width=1\linewidth]{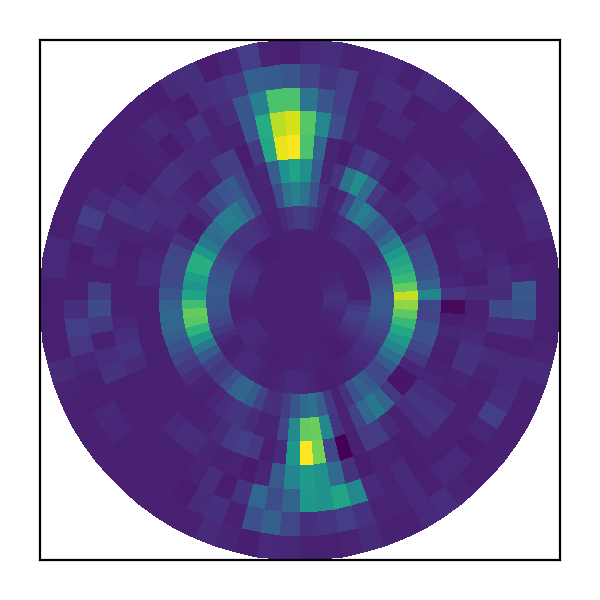}\\[-0.0em]
		\footnotesize (b)
	\end{minipage}
        \\
	\begin{minipage}{0.3\linewidth}
		\centering \includegraphics[width=1\linewidth]{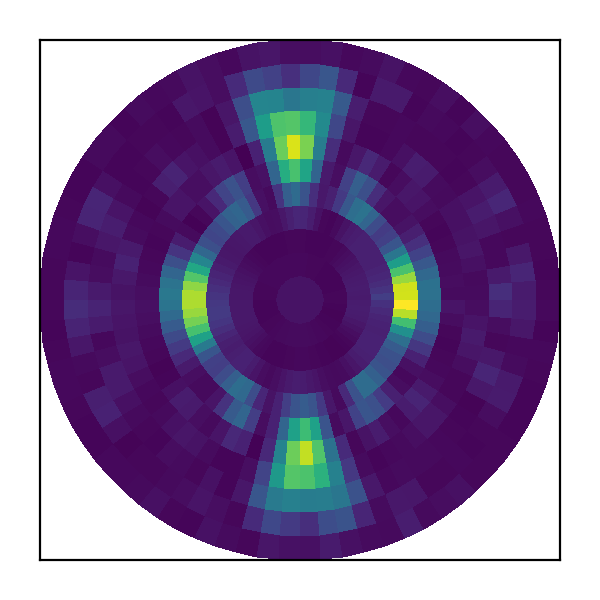}\\[-0.0em]
		\footnotesize (c)
	\end{minipage}
	\begin{minipage}{0.3\linewidth}
	\end{minipage}
	\begin{minipage}{0.3\linewidth}
		\centering \includegraphics[width=1\linewidth]{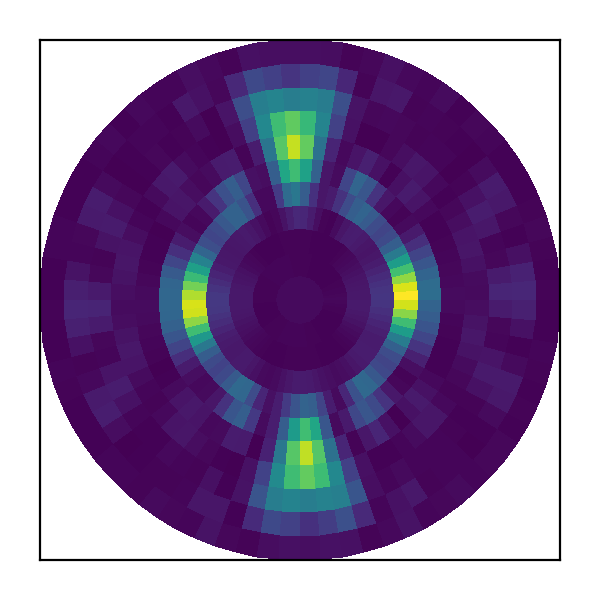}\\[-0.0em]
		\footnotesize (d)
	\end{minipage}
        \\
	\begin{minipage}{0.7\linewidth}
		\centering \includegraphics[width=1\linewidth]{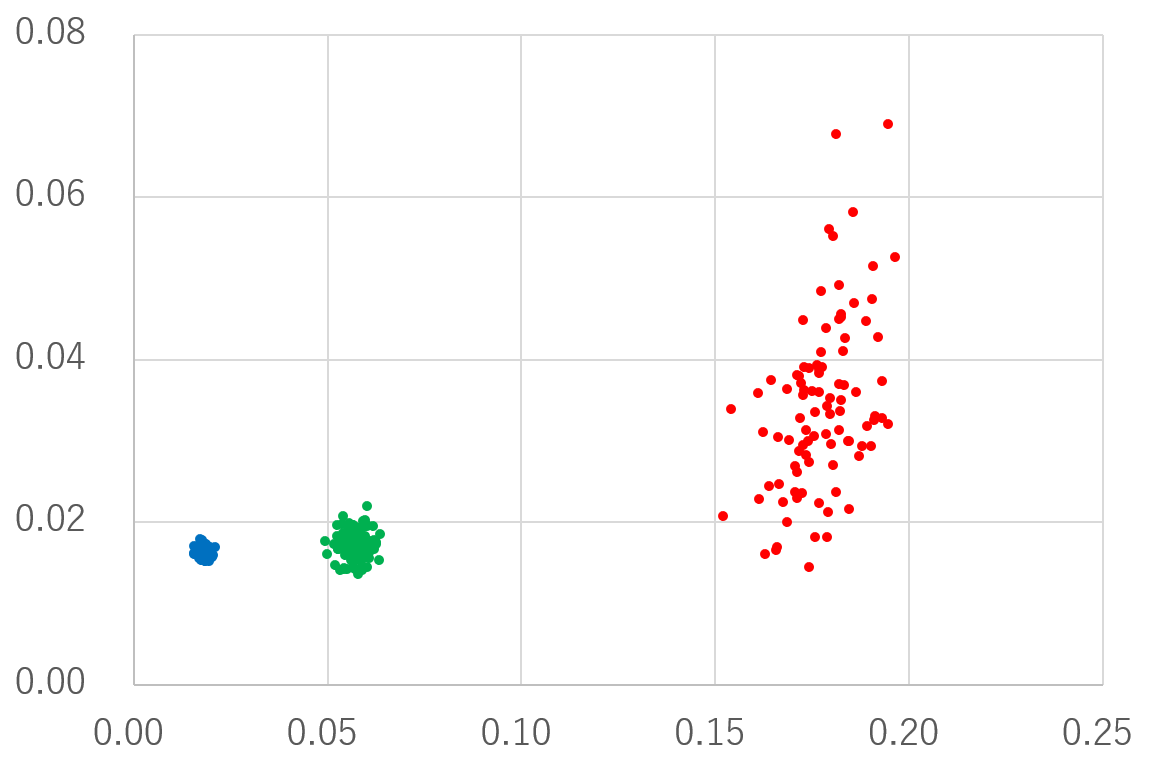}\\[-0.0em]
		\footnotesize (e)
	\end{minipage}
	\caption{(a) PMM without noise. 
		(b) -- (d) quasi PMM with noise added during total photoelectron number 
		$T = 10^3$, $10^4$, and $10^5$, respectively. 
		(e) MO-error as a function of total variation distance.  
	}
	\label{fig:sample-of-noisy-measurement}
\end{figure}

\subsection{Structure discrimination}
\par
This section shows that our method can identify the molecular structure from a single PMM 
\par
\begin{figure}[t]
	\centering
	\centering

        \begin{minipage}{0.2\linewidth}
		\centering \includegraphics[width=1\linewidth]
		{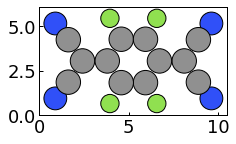}\\[-0.4em]
		\footnotesize (a) \\[0.4em]
        \end{minipage}       
        \begin{minipage}{0.2\linewidth}
		\centering \includegraphics[width=1\linewidth]
		{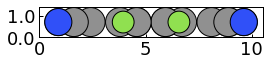}\\[-0.4em]
		\footnotesize (b) \\[0.4em]
        \end{minipage}         
        \begin{minipage}{0.2\linewidth}
		\centering \includegraphics[width=1\linewidth]
		{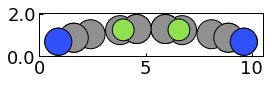}\\[-0.4em]
		\footnotesize (c) \\[0.4em]
        \end{minipage}
        \begin{minipage}{0.2\linewidth}
		\centering \includegraphics[width=1\linewidth]
		{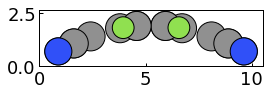}\\[-0.4em]
		\footnotesize (d) 
	\end{minipage} ~~
	\begin{minipage}{0.75\linewidth}
		\centering \includegraphics[width=1\linewidth]
		{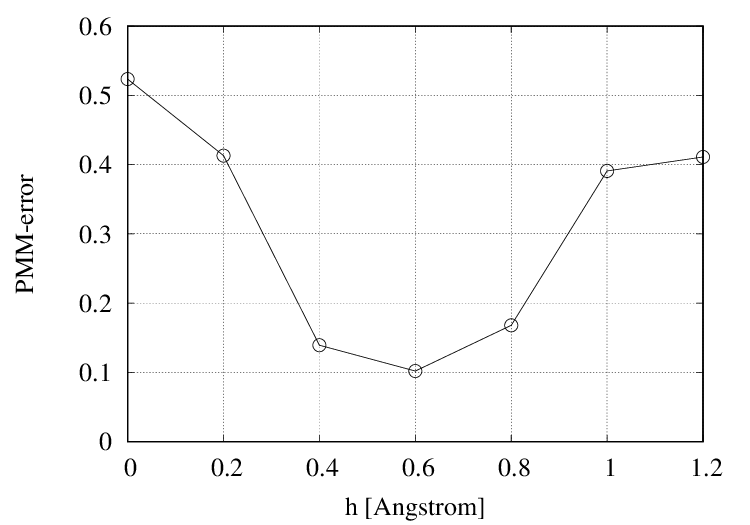}\\[-0.0em]
		\footnotesize (e)
	\end{minipage}
 	\begin{minipage}{0.75\linewidth}
		\centering \includegraphics[width=1\linewidth]
		{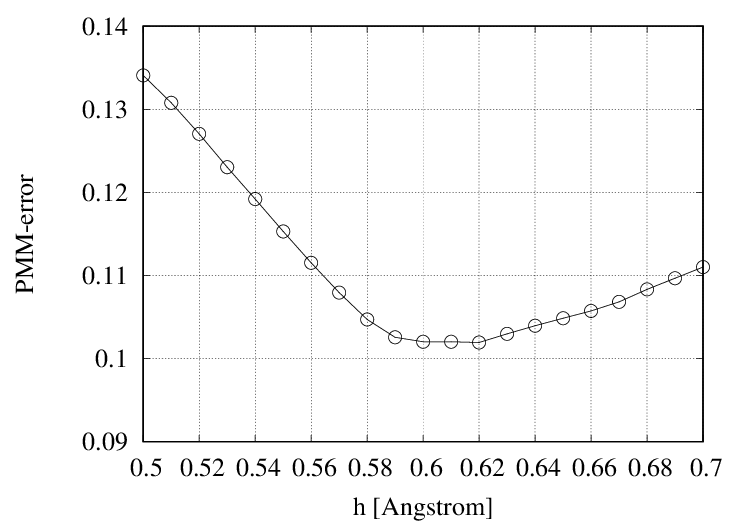}\\[-0.0em]
		\footnotesize (f)
	\end{minipage}
	\caption{
		Example of structure identification. 
		When the model structure coincides with the true structure that gives the PMM, the estimated PMM can become closer to the given PMM. 
		(a) F4-TCNQ, which is examined as an example of structure identification. 
		(b) Flat structure with height of $h = 0.0 $[\AA]. 
		(c) Bend structure with height of $h = 0.6 $[\AA], 
		that is used as a true structure. 
		(d) Bend structure with height of $h = 1.2 $[\AA]. 
		(e) Dependence of the PMM-error on the height of the model structure (in steps of 0.2 [\AA]) . 
            (f) Dependence of the PMM-error on the height of the model structure (in steps of 0.01 [\AA]). 
	} 
	\label{fig:structure-identification}
\end{figure}

We selected the F4-TCNQ molecule as an example, in which molecular orbitals increase the metal work function through an adsorption-induced geometric distortion of the F4-TCNQ on Cu(111) \cite{F4TCNQ}. During adsorption, the F4-TCNQ molecule is also bent without the rearrangement of atoms, in contrast to other adsorbed molecules. Figure \ref{fig:structure-identification} (a) shows the top view of F4-TCNQ. Figures \ref{fig:structure-identification} (b) - (d) show its flat and bent structures. We assumed that a quadratic function could represent the height of each atom. The height of the bent structure ($h$ [\AA]) is defined by the difference in height direction between the highest and lowest atoms. We first consider the following theoretical PMM $\vec{z}\,^{true}$: 
\begin{linenomath}
\begin{align}
	\vec{z}\,^{true} = \Delta^2 | A^{true} \vec{c}\,^{true} \, |^2, 
	\label{eq:z^true}
\end{align}
\end{linenomath}
where $\vec{c}\,^{true}$ is a given molecular orbital coefficients which is used as a true value, and $A^{true}$ denotes its corresponding measurement matrix based on the true structure. In this scenario, this theoretical PMM is called a \textit{true PMM}. 
\par
Next, we prepare the measurement matrix $A^{model}$ based on the model structures using (\ref{eq:A}). Lastly, we obtain the estimated molecular orbital coefficients $\vec{c}\,^{model}$ by solving 
\begin{linenomath}
\begin{alignat}{3}
	& \text{min} && \lambda_{tr} \mathrm{tr}(C) + \lambda_{norm} \|C\|_1 \notag\\
	& && + \sum_{m=1}^M | z^{true}_m - \Delta^2 \mathrm{tr}( A^{model}_m C) | \notag\\
	& \text{subject~to~~~} && C \succeq O, 
	\label{eq:robust-sparse-phaselift_2}
\end{alignat}
\end{linenomath}
where $z^{true}_m$ and $A^{model}_m$ are the $m$th entry of the vector $\vec{z}\,^{true}$ and the $m$th row vector of the matrix $A^{model}$, respectively. Then, we can obtain the estimated PMM: 
\begin{linenomath}
\begin{align}
	\vec{z}\,^{model} = \Delta^2 | A^{model} \vec{c}\,^{model} \, |^2. 
	\label{eq:model-based-estimated-PMM}
\end{align}
\end{linenomath}
We then calculate the PMM-error between the true and the estimated PMMs $d_{TV}(\vec{z}\,^{true}, \vec{z}\,^{model})$ using (\ref{eq:TV}). By preparing some model structures and applying this method repeatedly, we determined the model structure that yielded a PMM that is the closest to the true PMM in terms of the PMM-error. 
\par
By applying this method to a bent structure of F4-TCNQ, we have confirmed that PMM-error is minimized when the model structure coincides with the actual structure. Figure \ref{fig:structure-identification} shows an example of structure identification. Figure 4(e) shows the PMM-error as a function of $h$ in steps of 0.2. The curve has one minimum at $h=0.6$ [\AA]. Figure 4(f) shows the PMM-error for  $h$ changed in steps of 0.01. The minimum point is observed around $h=0.6$ [\AA] (a minimum point is $h=0.62$ [\AA]). XSW can be used to determine the structures of the adsorption system with an accuracy of 0.2 [\AA] \cite{F4TCNQ}, whereas our method can identify molecular structures with about six times higher accuracy. 

We prepared two model structures of F4-TCNQ ($h=0.55$ and 0.60 [\AA]) on Cu(111) and calculated the charge distribution of the Cu substrate using the plane wave-based Vienna ab initio simulation package \cite{VASP}. We observed a slight difference in the charge distribution of the substrate between the two models. It suggests that a slight structural change of the adsorbed molecule (even on a scale of 0.05 [\AA]) yields a slight change in the electronic state of the substrate. This method of estimating the structure from PMM can be applied not only to the case when structual information is available but also when it is not available.

\subsection{Structure estimation}
\par
In the previous section, we assumed several possible molecular structural models and determined their structures. This section introduces the method for estimating molecular structure without any molecular structure model. 
\par
One of applications of structure inference is again to detect bend structures. To infer structure of molecule, we prepare various bases that correspond to different heights for each atom. The robust sparse PhaseLift can choose appropriate bases to recover a given PMM. We can infer the height of each atom using information on which bases are chosen for its atoms. 
\par
A simple way of preparing such bases is to use measurement matrices for flat structures of different heights. Let $A^{(h)} \in \mathbb{C}^{M \times L}$ be a measurement matrix for flat structure, namely, the heights of all atoms are set to be $h$ [\AA]. We first give some values $h_1, \cdots, h_n$ for choosing the heights. Then, we can define a measurement matrix $A_{lattice} \in \mathbb{C}^{M \times nL}$ by concatenating $A^{(h_1)}, \cdots, A^{(h_n)}$ as 
\begin{linenomath}
\begin{align}
	A_{lattice}=(A^{(h_1)} | \cdots | A^{(h_n)}). 
\end{align}
\end{linenomath}
It should be noted that molecules that differ only in height yield the same PMM because of the time-shifting property of the FT. Therefore, for atoms the height of which can be assumed to be known, only one basis is given to avoid such a trivial ambiguity. Even in this case, we can obtain the molecular orbital coefficients by using (\ref{eq:robust-sparse-phaselift}), where $A$ is replaced with $A_{lattice}$. Because the molecular orbital coefficients give the heights of bases used for recovering a given PMM, the height of atoms in the molecule can be estimated. 
\par
We examined the bend structure of F4-TCNQ. Let the height of a true bend F4-TCNQ be 0.2 [\AA]. Let $A^{(h)}$ be a measurement matrix for the flat structure of F4-TCNQ at height $h$. Let $H=$ $\{0.0,$ $0.1,$ $0.2,$ $0.3\}$ be a set of heights, and we set $A_{lattice}$ $=$ $(A^{(0.0)}$ $|$ $A^{(0.1)}$ $|$ $A^{(0.2)}$ $|$ $A^{(0.3)})$. Without loss of generality, we can assume the height of nitrogens in F4-TCNQ is zero. We therefore remove bases for nitrogens of heights 0.1, 0.2, and 0.3 from $A_{lattice}$. 
\par
In structure estimation, the true positions of the atoms are treated as unknown. Therefore, the positions of the prepared bases are different from those of their corresponding atoms. We estimate the height of each atom using the positions of several bases. To simplify calculations, the atom heights are estimated based on the maximum value of the squared molecular orbital coefficients. First, the molecular orbital coefficient vectors of HOMO and HOMO1 are calculated by using the above method from the theoretical PMMs for HOMO and HOMO1. Next, we choose the basis whose squared molecular orbital coefficient is maximum for HOMO for each atom. Note that when the second largest molecular orbital coefficient is close to the largest value, i.e., their difference is less than about 10\%, both bases with the first and the second largest coefficient are included in the estimation. For HOMO1, we choose the basis in the same way. Lastly, the estimated height of each atom is obtained as an average value of the heights of all selected bases in both HOMO and HOMO1. 
\par
As a result, the height $h$ of the C atom in the benzene ring is 0.17 [\AA] and the height of the F atom connected to the benzene ring is 0.15 [\AA]. The height of the C atom connected to the nitrogen atom cannot be estimated due to the small electron density.
The height of the remaining carbon was estimated to be 0.10 [\AA].
The estimates and errors (differences between the estimated and true values) are shown in Table \ref{table:structure-inference}.  

\begin{table}[t]
	\caption{Structure inference of F4-TCNQ.}
	\label{table:structure-inference}
	\centering
 	\begin{tabular}{ccc}
	\hline
	atom & estimates & errors \\
	& [\AA] & [\AA] \\
	\hline
	C & 0.17 & 0.016 \\
	F & 0.20 & 0.014 \\	
	C & 0.07 & 0.051 \\
	N & 0.00 & -- \\
	\hline
	\end{tabular}
\end{table}
\par
The accuracy is expected to be further improved by changing the photoelectron energy $\epsilon_{\vec{k}}$ and performing the calculation in multiple layers. This analysis can be further refined; using parameter studies, more layers for determination, and PMMs with different photoelectron.

\section{Conclusion}
\par
This paper proposes a new POT method based on the PhaseLift method. Our PhaseLift POT method provides 3D orbital and structural information from a single theoretical PMM. Further, we demonstrated that our method is robust against unavoidable noise.In contrast to conventional POT, which identifies molecular orbitals and their energies, our technique reveals the structure of the molecule with an accuracy of 0.05 [\AA]. After the modification of the measurement matrix $A$, our method showed the scalability of analysis of the motion of atoms in multilayer and multiple rotational domains of the molecules on the substrates. 
\par
Our method allows to simultaneously obtain electronic and structural states from an experimental PMM; it is an innovative tool for the quantum-mechanical-based understanding of inter-molecular and molecule-substrate interactions in molecular thin films, spintronic materials, and photocatalytic surfaces.

\begin{acknowledgements}
\par
The authors would like to thank Misa Nozaki for knowledge of representing molecular orbitals by linear combinations of the atomic orbital and writing a part of the computer program on setting the orbital. The authors are grateful for the financial support from the Grant-in-Aid for Scientific Research S (23H05492, KM) and Scientific Research C (23K03841, KM, 20K05643, KN) from the Japan Society for the Promotion of Science. 
\end{acknowledgements}

\vspace*{2em}
\hfil\hfil{\bf AUTHOR CONTRIBUTIONS}\hfil
\vspace*{1em}
\par
R.A., R.S., K.N., and K.M. carried out the calculations. All authors were involved in the discussion of the results. K.N. and K.M. wrote the paper with significant contributions from M.H.


\end{document}